\documentclass{aa}

\makeatletter
\def\ps@headings{%
  \def\@oddhead{}%
  \def\@evenhead{}%
  \def\@oddfoot{\hfil\thepage\hfil}%
  \def\@evenfoot{\hfil\thepage\hfil}%
}
\makeatother

\usepackage{graphicx}

\usepackage{bm}
\usepackage{txfonts}
\usepackage{float}

\usepackage[colorlinks=true,citecolor=blue,linkcolor=blue]{hyperref}

\usepackage{xcolor}
\begin{document}

   \title{The Nucleated Atomistic Grain Growth Simulator (NAGGS): application to the size-dependent structural and physical properties of nanosilicate dust}

    \author{Joan Mariñoso Guiu
          \inst{1},  Antoni Macià Escatllar \inst{1}
          \and
           Stefan T. Bromley\inst{1,2}\thanks{Corresponding author: s.bromley@ub.edu}
          }

   \institute{Departament de Ciència de Materials i Química Física and Institut de Química Teòrica i Computacional (IQTCUB), Universitat de Barcelona, C/ Martí i Franquès 1, Barcelona, Spain.
         \and
            Institució Catalana de Recerca i Estudis Avançats (ICREA), Barcelona, Spain.\\
            \email{s.bromley@ub.edu}
             }

  \abstract
  {The essential physics of dust grains are typically rationalised using phenomenological approaches, often assuming highly simplified grain morphologies. Such descriptions are necessary if the underlying microscopic details are unknown or too complex to model. For small nanosized dust particles, these constraints can be overcome by atomistic simulations that can provide realistic detailed grain models.}
  {We show how atomistic forcefield-based simulations can be harnessed to: i) model the growth of structurally realistic nanograins, and ii) calculate a range of astrophysically relevant physicochemical properties directly from the growing grains.}
  {We report the Nucleated Atomistic Grain Growth Simulator (NAGGS) as a new tool to model the growth of realistic nanosized dust grains through the progressive accretion of monomers onto a nucleated seed. NAGGS can be used with open source molecular dynamics codes, allowing for the modelling of grains that have different chemical compositions and are grown under a range of astrophysical conditions.}
  {To demonstrate how NAGGS works, we use it to produce 40 nanosilicate grain models with diameters of $\sim$3.5 nm and consisting of $\sim$1500 atoms. We consider Mg-rich olivinic and pyroxenic grains, and growth under two circumstellar dust-producing conditions. We calculate  properties from the atomistically detailed nanograin structures (e.g. morphology, surface area, density, dipole moments) with respect to the size, chemical composition, and growth temperature of the grains.}  
  {Our simulations reveal detailed new insights into the complex interacting degrees of freedom during grain growth and how they affect the resultant physicochemical properties. For example, we find that surface roughness depends on the Mg:Si ratio during growth. We also find that nanosilicates have very high dipole moments, which depend on the growth temperature. Such findings could have important consequences (e.g. astrochemistry, microwave emission). In summary, our bottom-up physically motivated approach offers a detailed understanding of nanograins that could help in both interpreting observations and improving dust models.}

   
   \titlerunning{NAGGS}
   
   \maketitle

\section{Introduction}

Cosmic dust grains with sizes ranging from nanometres to micrometres constitute a small proportion of the mass of the interstellar medium (ISM) but play a fundamental role in many astrophyical processes. Dust absorbs and re-radiates stellar light and interacts with electromagnetic fields and ionised environments. These mechanisms combine to influence star formation and galactic morphology (\cite{Draine_review}). The small size of grains also leads to a relatively high proportion of available surface for adsorbing gas phase species and promoting chemical reactions \citep{Potapov_review}. Despite their central role in many astrophysical and astrochemical phenomena, the detailed atomistic processes that govern their growth, structure, and physicochemical properties under different environments remain incompletely understood.

The growth of grains depends sensitively on local parameters, such as gas density, temperature, and chemical abundances, which can vary drastically between different astrophysical environments. Observations can provide indirect constraints on these parameters, while laboratory experiments can approximate the extreme and varied range of physical conditions. Due to the complexity and inherently multiscale character of grain growth, theoretical models tend to focus on one aspect of the process. Herein, we focus on providing a bottom-up approach in which grain growth and grain properties are derived from an explicit atomistic level description. For grains with sizes above 100 nm with several millions of atoms, atomistically detailed modelling starts to become intractable. For such a grain size regime, phenomenological numerical modelling approaches can be employed to estimate grain growth rates (e.g. \cite{Hirashita_2011, Zhukovska_2018, Priestly_2021}. For smaller nanosized particles, atomistic modelling has shown that their structures and physicochemical properties can strongly depend on size and chemical composition (\cite{Nanoclusters_nucleation_2010, Escatllar_2019}). For circumstellar environments, explicit atomistically detailed models of molecular-scale species potentially participating in nucleation pathways have been proposed based on accurate quantum chemical density functional theory (DFT) calculations (e.g. \cite{Goumans_Bromley_2012, Goumans_Bromley_2013}. The results of these calculations can then be used to underpin numerical models of nucleation and growth of larger particles (e.g. \cite{Gobrecht_2016}). Calculations using DFT can also be used to model local surface regions of larger grains by using slab models with a periodically repeated cell. Such studies are primarily used for modelling surface astrochemistry (e.g. \cite{Periodic_DFT_Rimola_2023}) but are less suitable for describing the complex long-range structure and associated properties of whole grains. The computationally expensive nature of electronically detailed DFT calculations has thus far tended to limit them to systems of up to $\sim$200-300 atoms.

To provide a highly detailed description of grain structure and growth for larger particles, one can use computationally more efficient atomistic modelling methods. Such approaches often employ parameterised forcefields (FFs) that can provide an accurate description of interatomic interactions, while avoiding explicit calculation of the full electronic structure of the system (\cite{MD_nanoparticle_growth_review_2019}). In an astronomical context, there are a few examples of using FFs to model the properties of grains with a few hundred atoms. Ice grain models with up to 1000 water molecules have been built from the sequential placement of H$_2$O molecules using a mixed approximate quantum mechanical and FF approach (\cite{ACO-FROST_2022}). A FF-based approach has also been used to estimate sticking coefficients of atomic and molecular species on a pre-built amorphous carbon nanoparticle model containing 512 carbon atoms (\cite{Carbon_sticking_coeff_2024}).         

Here, we report a bottom-up FF-based approach for explicitly simulating atomistic level grain growth under astrophysical conditions. Our Nucleated Atomistic Grain Growth Simulator (NAGGS) can be used for growing grains from a few tens of atoms up to a few thousand atoms. With such information, size-dependent structural, physical, and spectroscopic properties can be followed from the molecular scale to the nanoscale. NAGGS thus helps to provide unprecedented insights into realistically grown nanograins. We highlight an example of how the NAGGS code can be employed through application to simulating the circumstellar growth of nanosilicate grains while tracking a set of astrophysically relevant properties. Although we use this specific case study to demonstrate our approach, the NAGGS code can be applied to the growth of grains with arbitrary chemical compositions at any selected temperature. As such, NAGGS can be used to describe a wide range of astrophysical environments and grain types.

Silicate dust is found in a wide range of astrophysical environments (\cite{Henning_review_2010}) and is believed to be mainly formed in the circumstellar envelopes of oxygen-rich asymptotic giant branch (AGB) stars in high-temperature nucleated growth processes (\cite{Gail_Sedlmayr_1998}). Assuming thermodynamic equilibrium and reasonable elemental abundances, it is predicted that the first circumstellar silicates to condense would be Mg-rich olivinic (i.e. Mg$_2$SiO$_4$) grains (\cite{Gail_Sedlmayr_1999}, \cite{Ferrarotti_2002}, \cite{Ferrarotti_2006}). Attempts to describe how such silicate grains form and grow have tended to focus on homogeneous nucleation of SiO monomers using classical nucleation theory (CNT) based approaches (\cite{Nuth_2006}, \cite{Paquette_2011}, \cite{Gail_2013}). CNT is a phenomenological equilibrium approach that employs parameters related to uniform macroscopic droplets, even down to the scale of nanoclusters. Such assumptions have led to doubts about the degree of applicability of CNT to nucleation of silicate dust grains (\cite{Donn_1985}, \cite{Cherchneff_2010} \cite{Paquette_2011-2}). Although clearly useful, CNT does not provide an atomistically detailed description of the structural and dynamical aspects of nucleation and growth processes. Without access to the detailed physical and chemical properties of growing grains, the degree to which CNT-based predictions can be tested observationally is inherently limited. 

Studies using CNT-based approaches tend to predict that significant circumstellar SiO nucleation should start at 600 - 900 K (\cite{Gail_Sedlmayr_1999}, \cite{Ferrarotti_2002}, \cite{Ferrarotti_2006}). Kinetic nucleation theory (KNT) calculations, which also employ accurate data from atomistic modelling, confirm that circumstellar nucleation of SiO monomers would occur only at the lower end of the temperature range predicted by CNT (i.e. close to 600 K) \citep{Bromley_2016}. However, the observed temperatures at the inner edge of silicate-forming dust shells are often found in the 900 - 1200 K range (\cite{Groenewegen_2009} \cite{Danchi_1994}). This strongly suggests that silicate grain formation and growth proceeds by processes that are more efficient than only SiO condensation at higher temperatures. For typical physical conditions in a circumstellar silicate dust condensation zone (i.e. 1200 – 1000 K, 0.1 – 0.001 Pa), O is locked in $\mathrm{H_{2}O}$ and SiO, while Mg is atomic (\cite{Gail_Sedlmayr_1998, Gail_Sedlmayr_1999}). By explicitly calculating the atomistic structures and corresponding thermodynamic stabilities of the clusters that could form from these monomers under such conditions, a viable grain growth route was proposed with alternating steps of oxidation and metal incorporation (\cite{Goumans_Bromley_2012}). This process has been implemented in KNT grain growth models and found to be in good agreement with observation (\cite{Sarangi_2013} \cite{Gobrecht_2016}). Other atomistic modelling work has focused on the issue of nucleation of the very first seeds to initiate silicate growth. Although silicates are highly refractory materials, in extreme circumstellar environments homogeneous nucleation is strongly thermodynamically hindered. Studies have thus investigated several small clusters (<50 atoms) of more refractory materials (e.g. TiO$_2$ and Al$_2$O$_3$ based species) as potentially viable nucleation seeds for silicates (\cite{Goumans2013}, \cite{Gobrecht_2022}, \cite{Gobrecht_2023}).

Unlike atomistic modelling attempts to understand the initial steps of silicate nucleation, NAGGS uses pre-nucleated seed particles and models the explicit addition of selected monomers at a chosen temperature. For the current nanosilicate growth case study, we employ an olivinic seed particle containing a few tens of atoms. From this initial seed structure, grain growth is followed for a sequence of monomer additions. This procedure currently allows us to access grains with a few thousand atoms having diameters of a few nanometres.

Ultrasmall nanosilicate clusters with fewer than 100 atoms and radii less than 0.5 nm have structural and spectral properties that are strongly affected by their size (\cite{Escatllar_2019} \cite{FaradayDiscuss_paper} \cite{Bromley_2024}). With increasing size, nanosilicate grains will increasingly exhibit properties associated with macroscopic laboratory silicate samples. However, it is not clear how quickly this transtion occurs with respect to grain size. Herein, we focus on nanograins with radii between 0.5 nm and 1.7 nm. Nanosilicate grains with radii up to 1.5 nm could constitute up to ten percent of the silicate grain mass (\cite{Aigen_2001}, \cite{AME_no_PAHS_Hensley_2017}). However, due to their small size, this relatively low mass fraction would imply a huge numerical abundance of such nanosilicates. The main evidence for the widespread abundance of such nanosilicates comes from their potential role as carriers of the anomalous microwave emission (see \cite{AME_no_PAHS_Hensley_2017, Hoang_2016, ame_toni}). Nanosilicates have also been reported to play a potentially important role in growth and nucleation processes in different astrophysical environments (e.g. circumstellar grain growth (\cite{Nuth_NanoSiOx_2022}), silicate cloud formation in exoplanets (\cite{Silicate_clouds_Nature2024}) and brown dwarfs (\cite{Silicate_clouds_brown_dwarfs_MNRAS2022}).
To follow the nanosilicate grain growth process in detail, we calculated a range of astrophysically relevant physicochemical grain properties (e.g. structure, morphology, density, dipole moments) and analyse how they depend on size, temperature, and chemical composition. The corresponding treatment of infrared spectra will be the focus of a follow-up study.

\section{The NAGGS code} \label{Code_functioning}
The primary objective of the NAGGS code is to provide a general platform to follow the atomistically detailed growth of dust grains from nucleated seed particles. For the purposes of this work we refer to grain growth as a process of accretion of gas phase species. In future versions of the code we also plan to include growth by grain-grain aggregation. NAGGS simulates the accretion process by controlling sequential collisions of monomers with the growing grain using atomistic molecular dynamics (MD) simulations. The NAGGS runs are initially set-up by specifiying: 1) the size, structure and chemical composition of the initial seed particle, 2) a library of monomers, \textit{i.e.} the atomic and molecular species that will be used to grow the grain, 3) the desired chemical composition of the grains (e.g. for magnesium silicates this could bias the stoichiometry to pyroxenic (Mg:Si ratio of 1) or olivinic (Mg:Si ratio of 2) grains), and 4) the temperature.
In this work, we model the growth of nanosilicate grains following the monomeric growth process proposed by \cite{Goumans_Bromley_2012}. In this scenario, if either Mg or SiO is introduced in one step, in the subsequent addition an oxygen atom is added, and vice versa. Note that we employ O monomers instead of OH or H$_2$O molecules following the simplifying and reasonable assumption that these latter species mainly act to add O to the growing particle and liberate gaseous H$_2$. The selection of either Mg or SiO monomers is dependent on a probabilistic bias of the growth towards the target chemical composition. Once an incoming monomer is chosen, its speed ($s$) is assigned from a Maxwell-Boltzmann distribution that is consistent with the specified temperature. The selected monomer is then placed at a random position around the grain at a radial distance from the centre of mass (COM) of the grain determined by 

\begin{equation}\label{distance_monomer}
    COM_{distance} = \bar{a} + (\alpha + \beta s),
\end{equation}

where $\bar{a}$ is the estimated radius of the grain (see below), $s$ is the monomer speed and $\alpha$ and $\beta$ are user-defined constants. The additive constant ensures that the incoming monomer starts sufficiently away from the grain surface, so that there are no significant monomer-grain interactions. The multiplicative constant allows for faster monomers to be placed further away from the surface which helps to normalise the number of MD steps for collisions of monomers with different speeds. To determine the velocity of each incoming monomer, a radial vector pointing from the monomer to the COM of the grain is first defined with a magnitude set to $s$. This vector is then displaced in a random perpendicular direction to the direction of the vector by a random radial offset (i.e. the impact parameter). In this work, the impact parameter range was set to be between zero and $0.9\bar{a}$.

Using this information, NAGGS generates the necessary inputs for the open access LAMMPS (\cite{lammps}) code, which then performs the classical FF-based MD simulations to perform the collision of the monomer with the grain. After each MD run, the resulting new grain structure is equilibrated to the ambient temperature and then serves as the starting point for the next monomer addition cycle. Taking the number density of circumstellar monomers to be 10$^{10}$m$^{-3}$ and a temperature of 1100 K, the average interval between collisions with a grain with $\bar{a}$ = 1 nm is expected to be of the order of 10 hours. Thus, each collision between a monomer and the growing grain is assumed to be independent of other collisions. 
 
Each NAGGS monomer addition cycle is divided into two phases. First, the internal temperature of the grain is equilibrated to the target value using a thermostat under the canonical ensemble (NVT: constant number of atoms, volume and temperature). In the second phase, the incoming monomer is introduced using the computed position and velocity, and the simulation is run under the microcanonical ensemble (NVE: constant number of atoms, volume and energy). In this work, we found that a 1.5 ps duration was sufficient for the energy of the impacting monomer to dissipate throughout the grain upon and after the collision. For the equilibration of the grain temperature before a subsequent monomer addition, we used a period of 2.5 ps. The timescales for equilibration and dissipation are more sensitive for the initial stages of growth when the grain has fewer than a couple of hundred atoms, and each collision has a proportionally larger effect. In principle, this means that the equilibrium time can be progressively reduced For larger grains without significantly affecting the results obtained. Herein, we retain the same reported MD times for the full NAGGS run in all cases.

The FF used to describe the interactions between all ions in the growing nanosilcates (i.e. Si$^{4+}$, Mg$^{2+}$ and O$^{2-}$) includes carefully parameterised analytical terms for long-range electrostatics and polarisation and short-range attractive van der Waals interactions and Pauli repulsion. The FF is based on that used to model nanosilicates in \cite{Escatllar_2019}, but with a slightly improved parameterisation of the Mg-O interaction. The full list of FF parameters used in the reported NAGGS simulations are listed in Tables \ref{IP_Charges}-\ref{IP_Core-Shell} in Appendix A. To assess the capacity of this FF to describe relative energetic stabilities of nanosilicate structures, the set of 90 nanosilicate used in \cite{FaradayDiscuss_paper} was used as a reference test set. Each of the nanosilicate structures in this latter set was optimised using the FF used in the present work and with accurate quantum chemical DFT calculations. Figure \ref{dft_ff_relative_energies} in Appendix B shows the relative energy per formula unit obtained from each approach for pyroxene and olivine nanosilicates, showing a good correlation between both methods for each composition.
 We note that NAGGS can be used with a wide range of FFs (e.g. FFs parameterised for other elements, machine-learned FFs). In principle, this allows NAGGS to model the growth of many complex nanograin types with different astrophysically relevant morphologies and compositions (e.g. carbonaceous, silicate-carbonaceous, silicate-ice).

\begin{figure*}
\centering
\includegraphics[width=17cm]{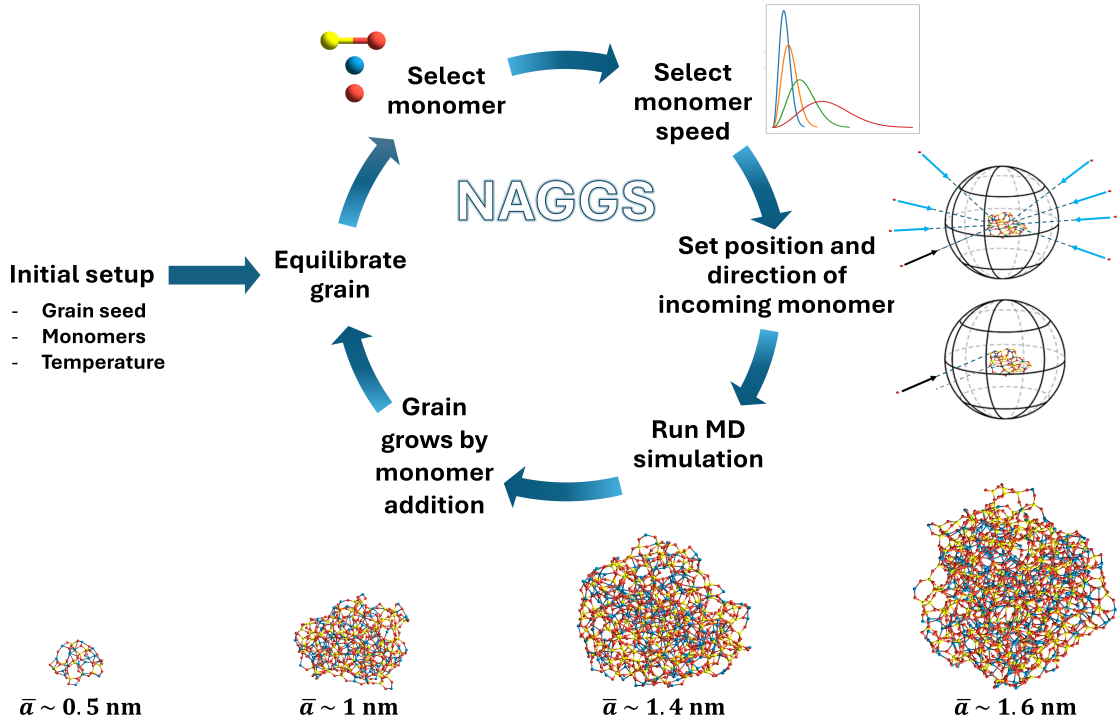}
  \caption{Upper: Summary of the steps involved in a NAGGS simulation. Lower: Nucleated growth of a nanosilicate grain with respect to the grain radius. Atom colour key: Mg - blue,  Si - yellow, O - red.}
     \label{nucleation_code_summary}
\end{figure*}

We note that the NAGGS set-up described above leads to almost all incoming monomers remaining on the nanosilicate grain surface close to the site of the collision in all simulations. This suggests that silicate grain growth by accretion could have a very high sticking coefficient. This efficient growth is likely driven by the strong directing interactions between the monomers and the ions in the growing grain. As neutral monomers get closer to the grain, they first become polarised by the ions in the grain, which drives an increased attraction. Eventually, when very close to the grain, the monomers are incorporated as ionic species through charge transfer with the low coordinated ions on the grain surface. The use of an ionic FF essentially assumes that the process of polarisation and charge transfer occurs earlier in the collision process. Further tests are required to see if this choice could affect the sticking coefficient. We stress that the aim of the present work is not to predict grain growth rates but to derive realistic nanograin structures from a physically motivated accretion process and analyse the properties of the resulting grains. An overview of how NAGGS operates is illustrated in Figure \ref{nucleation_code_summary}.
 
\begin{figure}[h]
\centering
\includegraphics[width=\hsize]{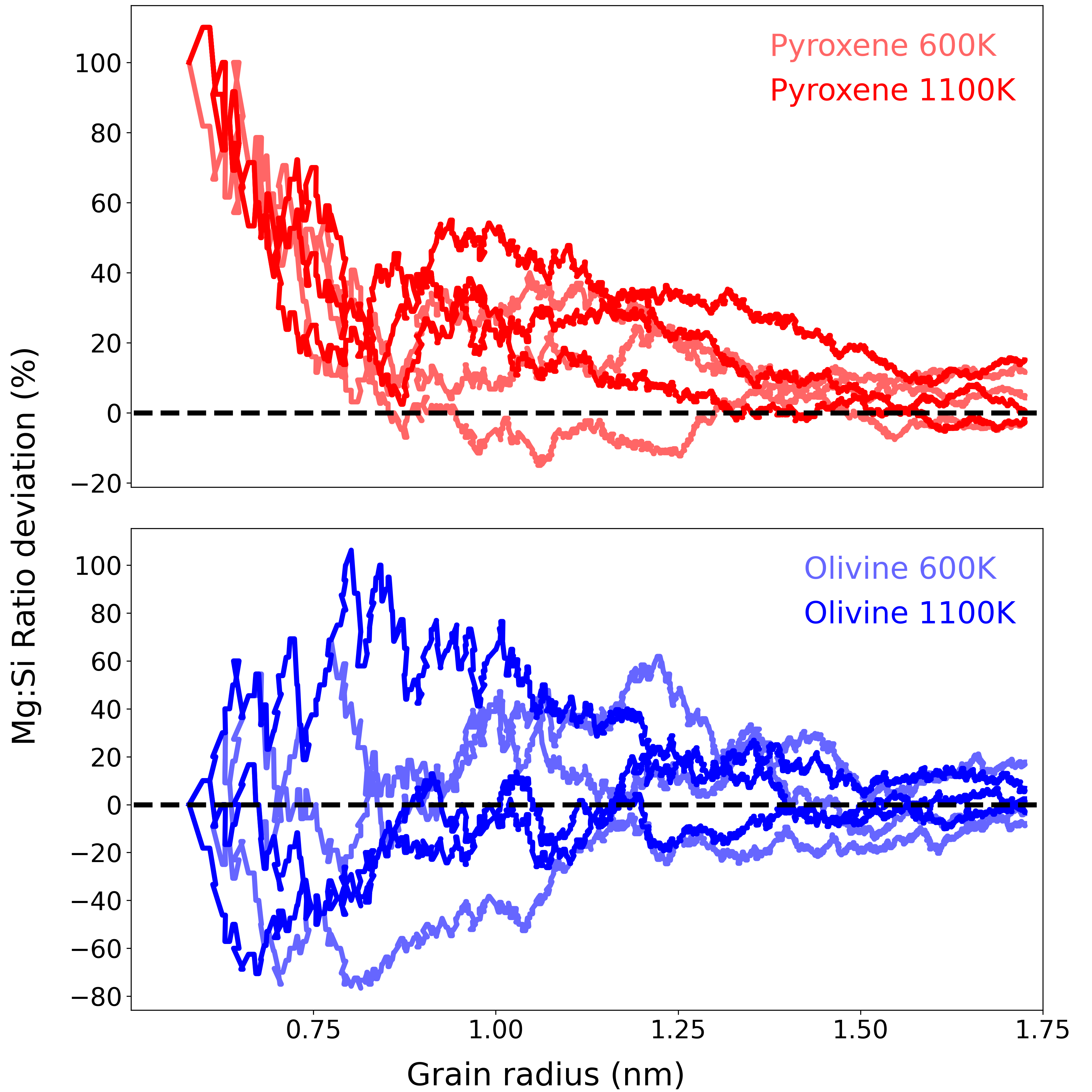}
  \caption{Size-dependent convergence of the Mg:Si ratio for pyroxenic (top) and olivinic (bottom) grain compositions, grown at 600 K and 1100 K. For each chemical composition and temperature, the results of three different NAGGS simulations are shown for each composition-temperature combination. The plots show percentage deviations of the Mg:Si ratio relative to the corresponding stoichiometric values (i.e. zero deviation highlighted by dashed lines). The same analysis for the O:Si ratio can be found in \ref{o_si_ratio_plot}.}
     \label{mg_si_ratio}
\end{figure}

\section{Nanosilicate grain properties}
We used NAGGS to model four different scenarios, from which we extracted grains of varying size and analysed them according to a range of physicochemical properties. To ensure that all NAGGS runs were independent, each run used different random seeds for generating the stochastic selections of incoming incoming monomers, speeds and impact parameters. Following the thermodynamic preference for olivine in high temperature circumstellar nucleation, for all NAGGS simulations a low energy 70 atom olivine cluster (Mg$_2$SiO$_4$)$_{10}$ (\cite{Escatllar_2019}) was used as an initial seed. The same seed structure was employed in all simulations to avoid any potential dependence of the growth process on initial conditions when comparing and averaging results from different runs. We considered the growth from this seed towards larger olivinic nanograins and towards nanograins that becomes progressively more pyroxenic. These two scenarios are compared for two temperatures, which correspond approximately to upper (1100 K) and lower (600 K) limits for circumstellar silicate grain growth (see the introduction). The four considered growth scenarios can be summarised as:    

\begin{itemize}  
  \item Mg-rich olivinic (Mg$_2$SiO$_4$) composition at 600K,
  \item Mg-rich olivinic (Mg$_2$SiO$_4$) composition at 1100K,
  \item Mg-rich pyroxenic (MgSiO$_3$) composition at 600K,
  \item Mg-rich pyroxenic (MgSiO$_3$) composition at 1100K.
\end{itemize}

We note that the NAGGS runs under these conditions produce amorphous silicate grains. Condensation models of circumstellar silicate dust growth predict that silicates usually condense initially in amorphous form. Subsequently the grains can become crystalline if they pass through sufficiently hot zones (>1100K) with sufficiently long residence times (\cite{Gail_Sedlmayr_1999}). From experiments on amorphous silicate samples, no crystallinity is induced at 600K. At 1100K we are in a transitional temperature range at which some structural annealing towards crystallinity can occur after several hours (\cite{Fabian_2000, Hallenbeck_2000}). However, observations of evolved-star envelopes, where temperatures close to the star can exceed 1100K, reveal that crystalline silicates usually comprise a minor fraction of the dust population (\cite{Crystalline_silicates_2002}). This indicates that timescales sufficient for high temperature crystallization of growing grains are infrequently met in these environments. We also note that modelling studies also show that amorphous silicates are intrinsically more stable than crystalline grains at the nanoscale size regime relevant to NAGGS simulations (\cite{Zamirri-NCs_2019, Escatllar_2019}).

For each of the four scenarios we performed ten separate NAGGS simulations, where each individual simulation consisted of 1200 monomer additions. All reported results for each analysed property were computed as a statistical average over all ten runs for the corresponding scenario. Below, we analyse a selection of data from the simulations.

\subsection{Chemical composition}

To demonstrate the ability of NAGGS to grow grains towards a desired chemical composition, Figure \ref{mg_si_ratio} shows the evolution of the deviation in the Mg:Si ratio from the corresponding target ratio for selected NAGGS runs for each simulated scenario. In the case of growing pyroxenic grains the initial Mg:Si ratio shows a large deviation from the target Mg:Si ratio of 1:1, as they originate from the olivinic seed with a Mg:Si ration of 2:1. However, as the NAGGS simulations progress, the chemical composition is progressively adjusted. After 1200 monomer additions, the average chemical composition is close to being pyroxenic, with a corresponding deviation from the target Mg:Si ratio of about 10 percent on average. On the other hand, the growth of olivinic grains starts from a seed with the appropriate target Mg:Si ratio. Here, we see statistical fluctuations of the ratio throughout, which decrease in magnitude as the NAGGS simulation proceeds. As for pyroxenic grain growth, the target composition for olivinic grains is reasonably well converged after 1200 monomer additions.

\subsection{Grain radius}

To estimate the average radius of the growing nanosilicate grains, we first extracted the three moments of inertia (i.e. $I_x, I_y, I_z$) of the grain at each NAGGS step directly from its atomistic structure and the atomic masses. We then represented the average shape of the grain as an ellipsoid with these three moments of inertia. The corresponding three radii of the ellipsoid are given by
\begin{equation}\label{Radius_x}
a_x = \sqrt{\frac{5}{2M}}(I_y+I_z-Ix),
\end{equation}
\begin{equation}\label{Radius_y}
a_y = \sqrt{\frac{5}{2M}}(I_x+I_z-Iy),
\end{equation}
\begin{equation}\label{Radius_z}
a_z = \sqrt{\frac{5}{2M}}(I_x+I_y-Iz),
\end{equation}
where ($M$) is the total mass of the grain. 

\begin{figure}[h]
\centering
\includegraphics[width=\hsize]{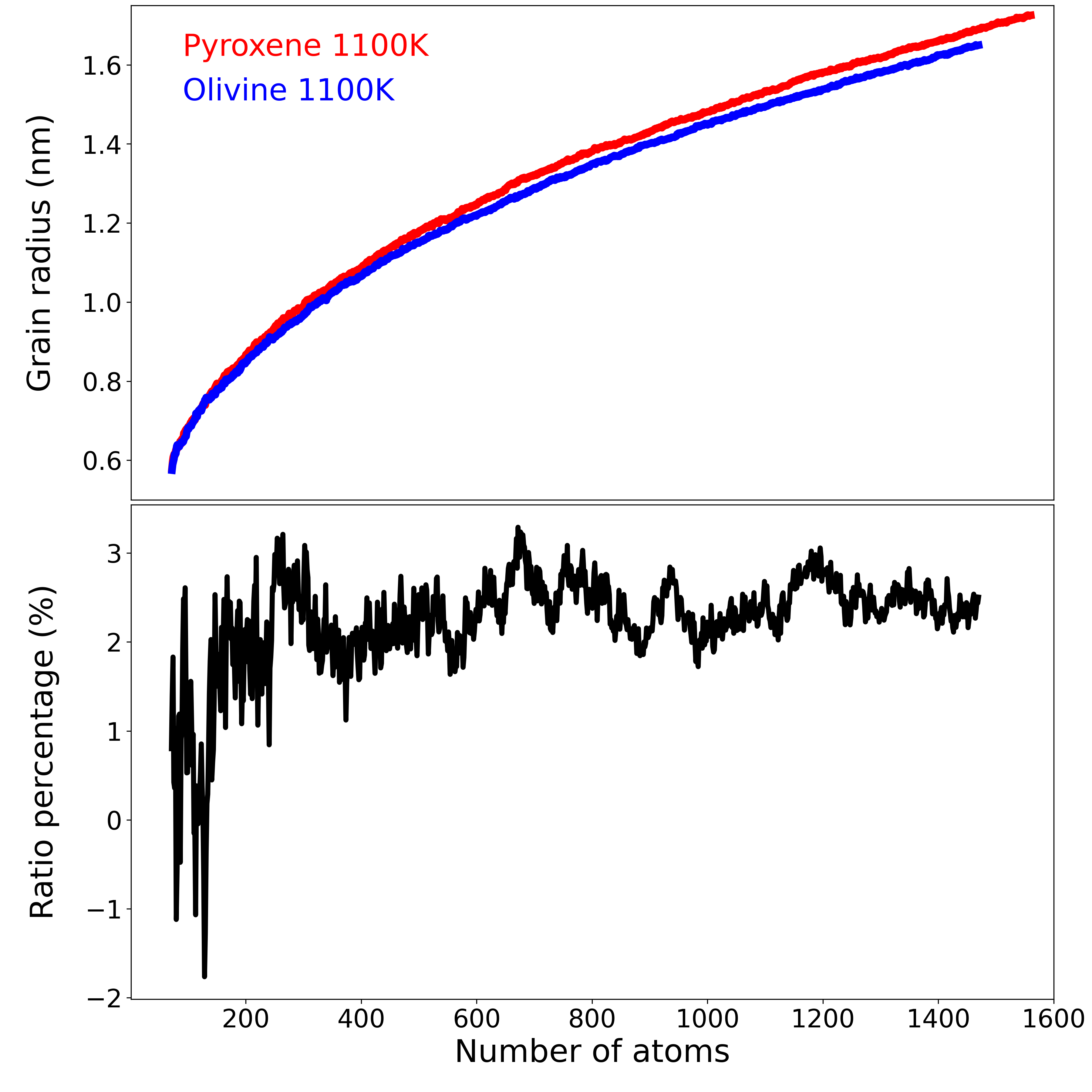}
  \caption{Evolution of nanosilicate grain radius (top) of pyroxene (red) and olivine (blue) grains at 1100K with respect to the number of atoms in the grain. Bottom panel shows the corresponding ratio (as a percentage) of the radii of pyroxene to olivine grains. The results for each grain composition were obtained from averages over ten NAGGS simulations. The corresponding analysis for grains grown at 600K can be found in \ref{radius_600K}.}
     \label{plot_radius}
\end{figure}

The estimated radius, $\bar{a}$, of the growing grain is then taken to be the mean value of these three radii. The evolution of $\bar{a}$ shows the range of sizes covered by the NAGGS simulations, ranging from particles containing only a few tens of atoms to larger particles with radii >1.5 nm and well over 1000 atoms (see Figure \ref{plot_radius}). This range includes those of molecular silicate species ($\bar{a} < 0.5$ nm) to small nanosilicate grains (\cite{Bromley_2024}). The results also show that after about 300 monomer additions olivinic nanograins tend to be between $1.5 - 3 \% $ smaller than pyroxenic nanograins for the same number of atoms. This indicates that olivinic nanograins possess more compact structures.   

\subsection{Grain sphericity}
Sphericity ($\Psi$) is a measure of how closely the shape of a particle approximates a perfect smooth sphere. This metric is defined as the ratio between the surface area of a sphere with the same volume as the object under study and the actual surface area of the object \citep{sphericity}:

\begin{equation}\label{sphericity_equation}
    \Psi = \frac{\pi^{1/3}(6V_{obj})^{2/3}}{A_{obj}},
\end{equation}

where $V_{obj}$ is the volume of the object and $A_{obj}$ its surface area. Sphericity values close to one indicate shapes close to being spherical, while lower values indicate non-spherical grain morphologies and/or increases surface roughness. To obtain the volume and surface area of the grains, we employed the MoloVol code \citep{molovol}, which computes the van der Waals volume and surface of atomic structures. In the calculation of both the volume and the surface, a probe of radius of 1.2~{\AA} and a grid resolution of 0.4~{\AA} were used. Radii of 2.19~{\AA}, 2.51~{\AA} and 1.5~{\AA} were considered for silicon, magnesium and oxygen atoms, respectively.

\begin{figure}[h]
\centering
\includegraphics[width=\hsize]{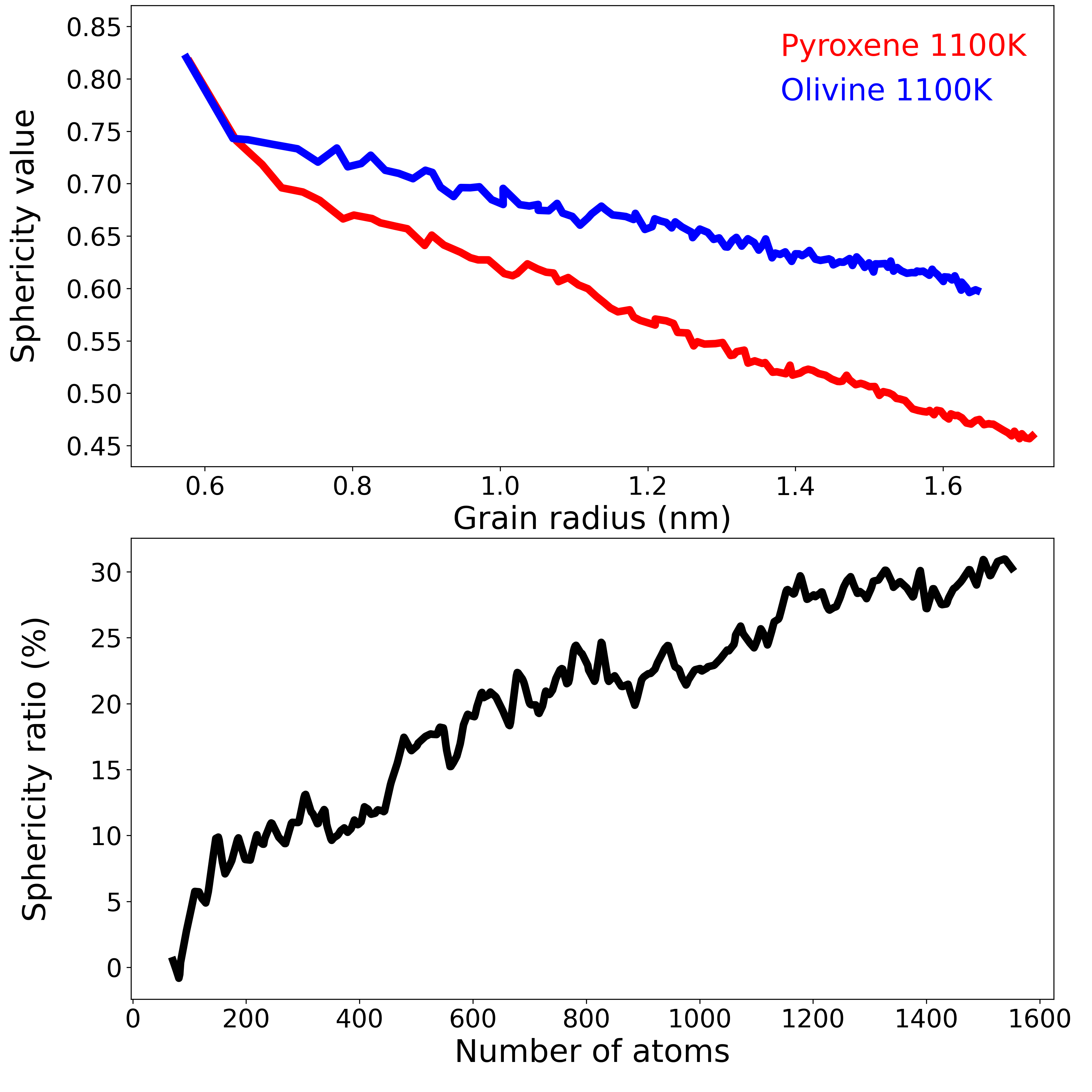}
  \caption{Top: Silicate grain sphericity evolution with respect to the number of atoms in the grain for both pyroxene (red) and olivine (blue) composition obtained at 1100K. The bottom panel shows the corresponding ratio (as a percentage) of pyroxene to olivine sphericity. The results were obtained from averages over ten NAGGS simulations. The corresponding analysis for grains grown at 600K can be found in \ref{sphericity_600K}.}
     \label{plot_sphericity}
\end{figure}

The results presented in Figure \ref{plot_sphericity} clearly show that the sphericity of the grains tends to decrease as they grow. This tendency is especially apparent for pyroxenic grains. Because the NAGGS code adds incoming monomers in a stochastic and statistically isotropic way, any decrease in sphericity is likely due to the heterogeneous set of interactions between the different ions in the grain. For example, the Si$^{4+}$ and Mg$^{2+}$ cations strongly repel each other, while both have strong attractive interactions with O$^{2-}$. While the thermodynamically favoured product of a stoichiometric mix these species would be a magnesium silicate, during stochastic growth local kinetic trapping can produce non-mixed regions that can accumulate and persist (e.g. segregated MgO or SiO$_x$).

\subsection{Grain morphology}
To better understand the origin of the decrease in the sphericity of the growing grains, we analysed the morphology of the grains independently of the surface roughness. To achieve this, we used the inertial moments of the growing grains to classify the them as prolate ($I_z = I_y > I_x$), oblate ($I_z > I_y = I_X$) or spherical ($I_x = I_y = I_z$). In cases where all three inertia moments differ, the grain shape lies between a prolate and an oblate spheroid. Strictly speaking, this classification is not purely geometric, as it also considers the distribution of mass throughout the grain. Although we do find some atomic segregation in the grains (see below), the average spatial distribution of atom types throughout the whole grain is found to be sufficiently regular as not to significantly affect this classification.

\begin{figure}
\centering
\includegraphics[width=9cm]{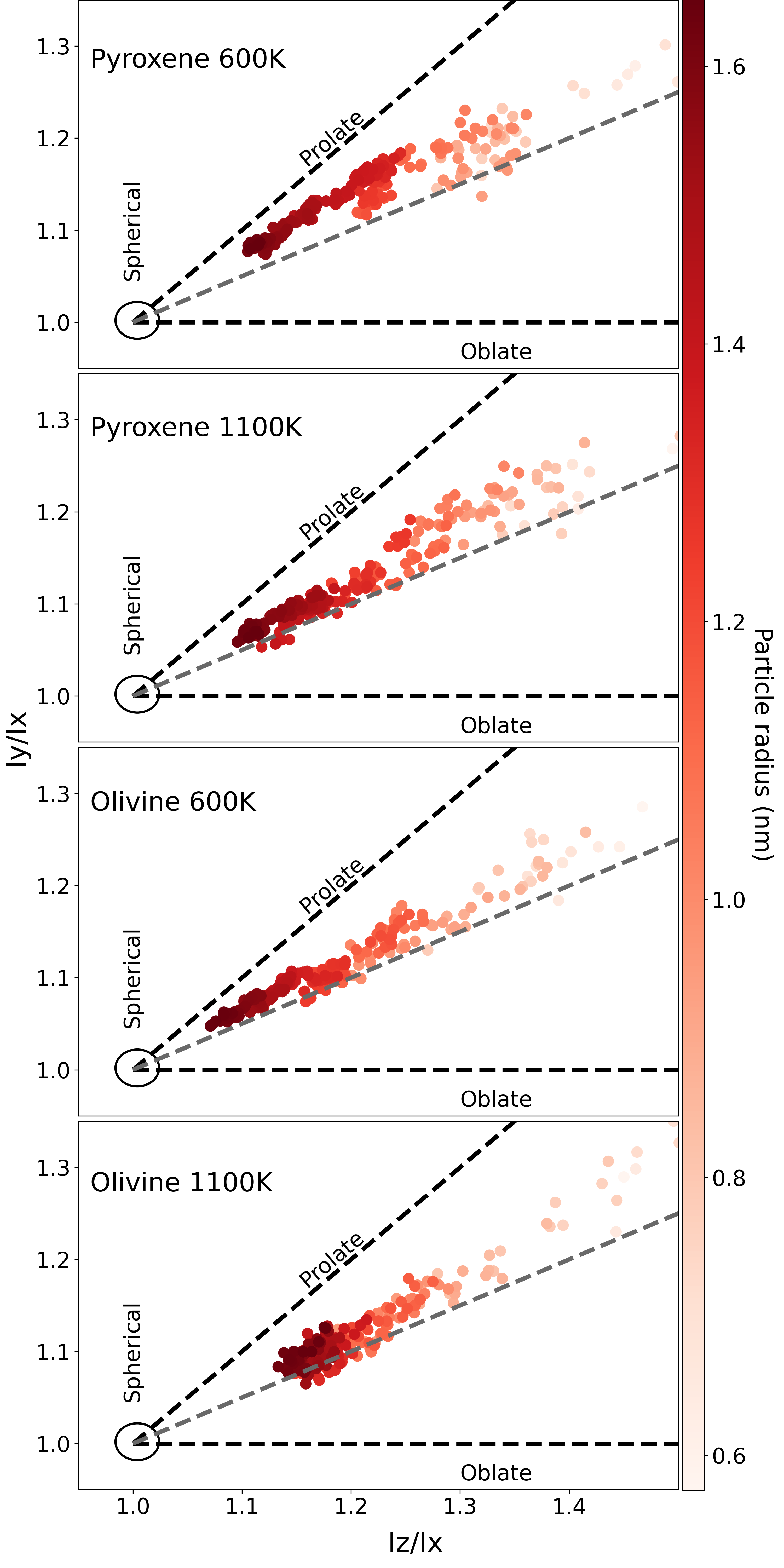}
\caption{Silicate grain shape evolution for the simulation of pyroxenic composition at 600K  pyroxenic composition at 1100K , olivinic composition at 600K and olivinic composition at 1100K (from top to bottom). The black circle indicates the ratio defining a perfect sphere. The grey dashed line corresponds to the cases in which grains would be equally prolate and oblate. The results were obtained as the average over ten NAGGS runs.}
     \label{plot_inertia}
\end{figure}

The evolution of the grain shape with respect to grain size is shown in Figure \ref{plot_inertia}. During the considered growth regime, all grains maintain a prolate tendency. The slight preference for a prolate grain shape appears to be relatively independent of the chemical composition and growth temperature. However, the main overall trend is that nanosilicate grains tend to become more spherical with increasing size. Extrapolating this latter trend to larger grain sizes, we would expect circumstellar grains grown from accretion to be spherical - as classified by their moments of inertia. As the average grain shape cannot explain the significant differences in sphericity shown in Figure \ref{plot_sphericity}, we turn our attention to the surfaces of the grains.   

\subsection{Grain surface area} 

\begin{figure}
\centering
\includegraphics[width=\hsize]{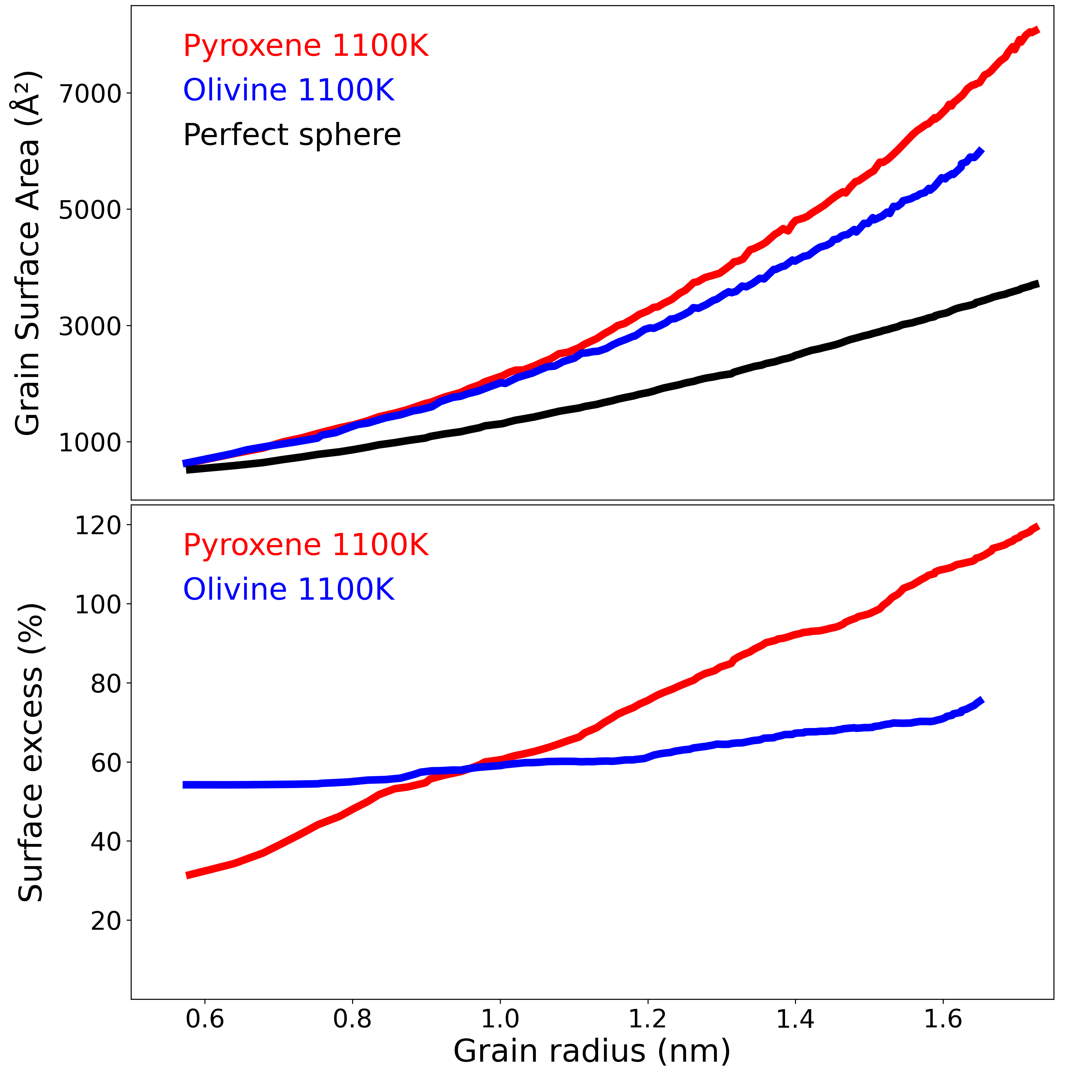}
  \caption{Top: Surface area of growing pyroxenic (red) and olivine (blue) grains grown at 1100K, and of a sphere with same radius as the grain (black). Bottom:  Percentage of excess surface area of the grains with respect to the corresponding sphere. The results were obtained from averages over ten NAGGS simulations. The corresponding analysis for grains grown at 600K can be found in \ref{surface_600K}.}
     \label{surface_increase}
\end{figure}

\begin{figure}[h]
\centering
\includegraphics[width=\hsize]{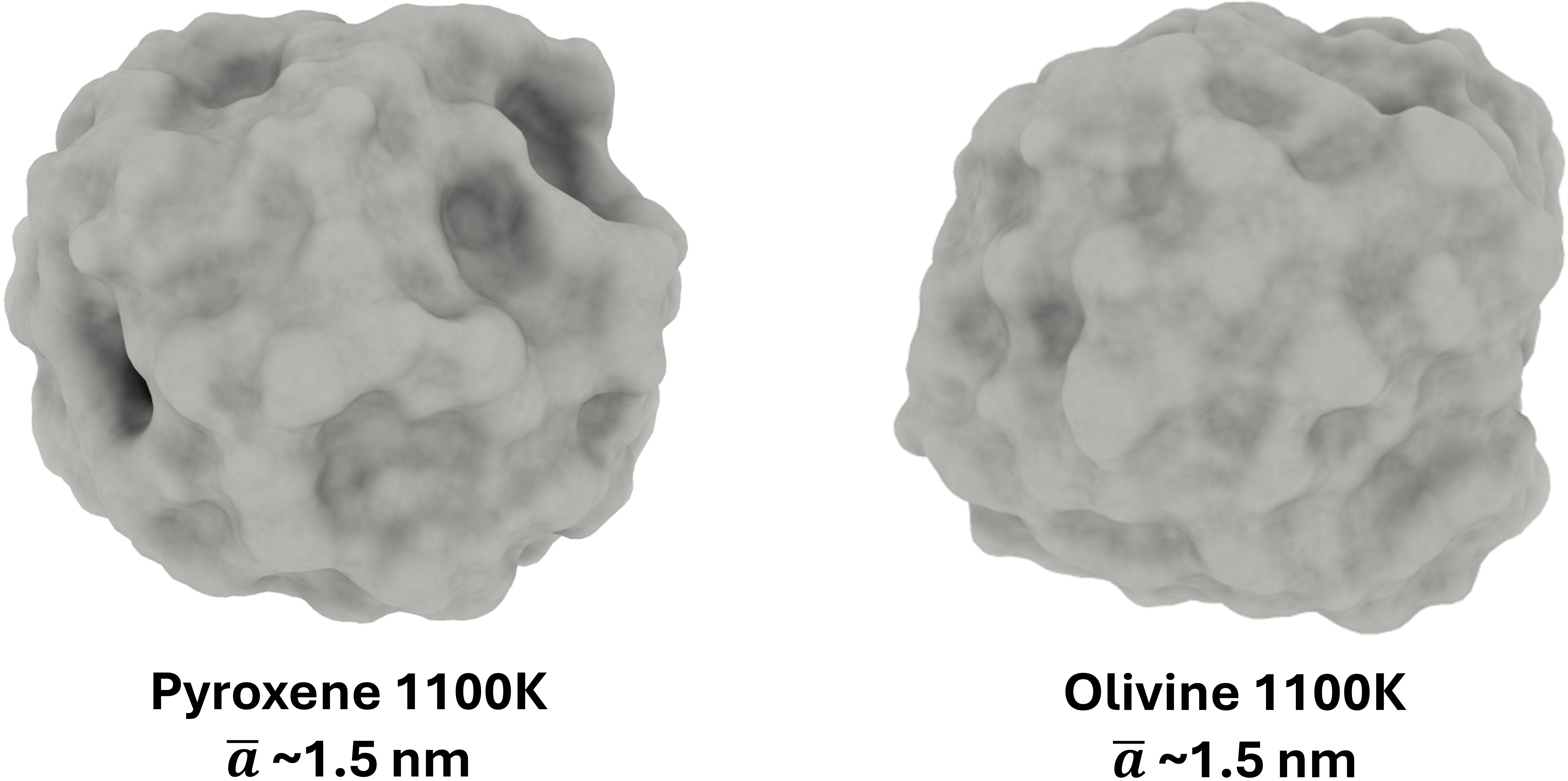}
  \caption{Visual representation of the surfaces of two typical grains with pyroxene (left) and olivine (right) composition. Both grains were obtained by NAGGS simulations at 1100K and have $\sim$1.5 nm radii.}
     \label{surface_roughness}
\end{figure}

If the grains become more spherical as they grow, the decrease in sphericity observed in Figure \ref{plot_sphericity} should be attributed to an increase in surface roughness. Visualization of the surface of a typical olivinic and pyroxenic nanograin, as scanned using a virtual atomic probe \citep{chimerax} (see Figure \ref{surface_roughness}), clearly supports this conclusion. For both grains, the overall shape is quite spherical; however, the pyroxenic nanograin clearly exhibits a considerably higher degree of surface roughness. To further quantify this difference, we compared the surface area of the olivinic and pyroxenic grains throughout their growth with that of a perfect sphere having a correspondingly equal volume. This analysis (see Figure \ref{surface_increase}) clearly shows a significant percentage excess of surface area for both types of grain. The surface excess of olivine grains is found to be $\sim$55\% at the start of the growth simulations, and increases gradually with size. The surface excess of pyroxene grains starts at a value lower than that of olivinic grains ($\sim$30\%) but increases more rapidly with increasing size. At the largest sizes reached at the end of the NAGGS runs, olivine grains have around 75\% excess surface area, whereas pyroxene grains have over 120\% excess surface area. 

The higher sphericity of pyroxene nanograins compared to olivine nanograins also has implications for their relative specific surface area (i.e. the surface area per mass). The specific surface area determines the proportion of available sites for interactions with gas phase species (e.g. adsorption and chemical reactions) for a given dust population mass. For both nanograins compositions we find specific surface areas of over one million m$^2$/kg, in line with their extreme small size. For the smallest sizes the nanograins have their largest specific surface area $\sim$ 2.7 $\times$ 10$^6$ m$^2$/kg, which then decreases with increasing grain size. The relative rate of decrease in the specific surface area for pyroxenic grains is clearly slower than for olivinic grains, which can be attributed to their lower sphericity.       

We note that the results in this sub-section are found to be relatively independent of the growth temperature (see the appendices). Overall, these tendencies suggest that pyroxenic nanograins would have distinct astrochemical properties to olivinic nanograins, which could also persist in larger grains.

\begin{figure}
\centering
\includegraphics[width=\hsize]{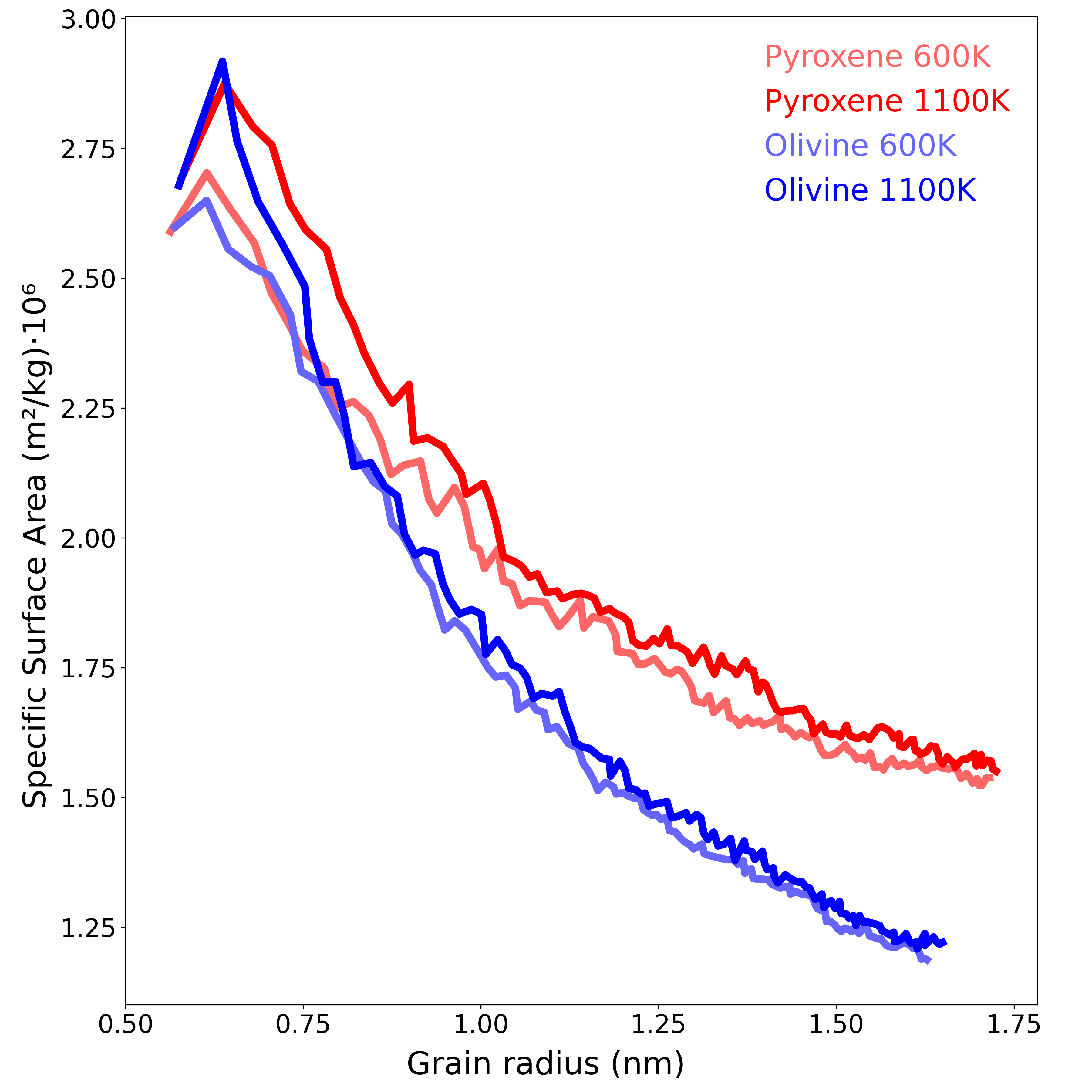}
  \caption{Specific surface area evolution for the simulation of pyroxenic (red) and olivinic (blue) grains grown at 600 K and 1100 K. The results were obtained from averages over ten NAGGS simulations.}
     \label{specific_surface}
\end{figure}

\subsection{Silica segregation}

Mg-rich silicates can be represented stoichiometrically as combinations of MgO (i.e. magnesia) and SiO\textsubscript{2} (i.e. silica). In enstatite crystals, the 1:1 (SiO\textsubscript{2})(MgO) combination forms oxygen-sharing linear chains of SiO\textsubscript{4}\textsuperscript{4-} tetrahedra, with Mg\textsuperscript{2+} occupying the space between the chains. In contrast, in the relatively silica-poor (SiO\textsubscript{2})(MgO)\textsubscript{2} stoichiometry of forsterite crystals, the SiO\textsubscript{4}\textsuperscript{4-} tetrahedra are isolated, with Mg\textsuperscript{2+} ions filling the space between them. We note that crystalline grains are not thermodynamically favoured for small nanograin sizes \citep{IR_and_crystallinity}. In all our NAGGS runs the nanosilicate grains grow by stochastic accretion and are intrinsically amorphous. In such nanosized systems it is not clear if the tendencies for silica segregation will follow those of the corresponding bulk crystalline silicate phase.

\begin{figure}[h]
\centering
\includegraphics[width=\hsize]{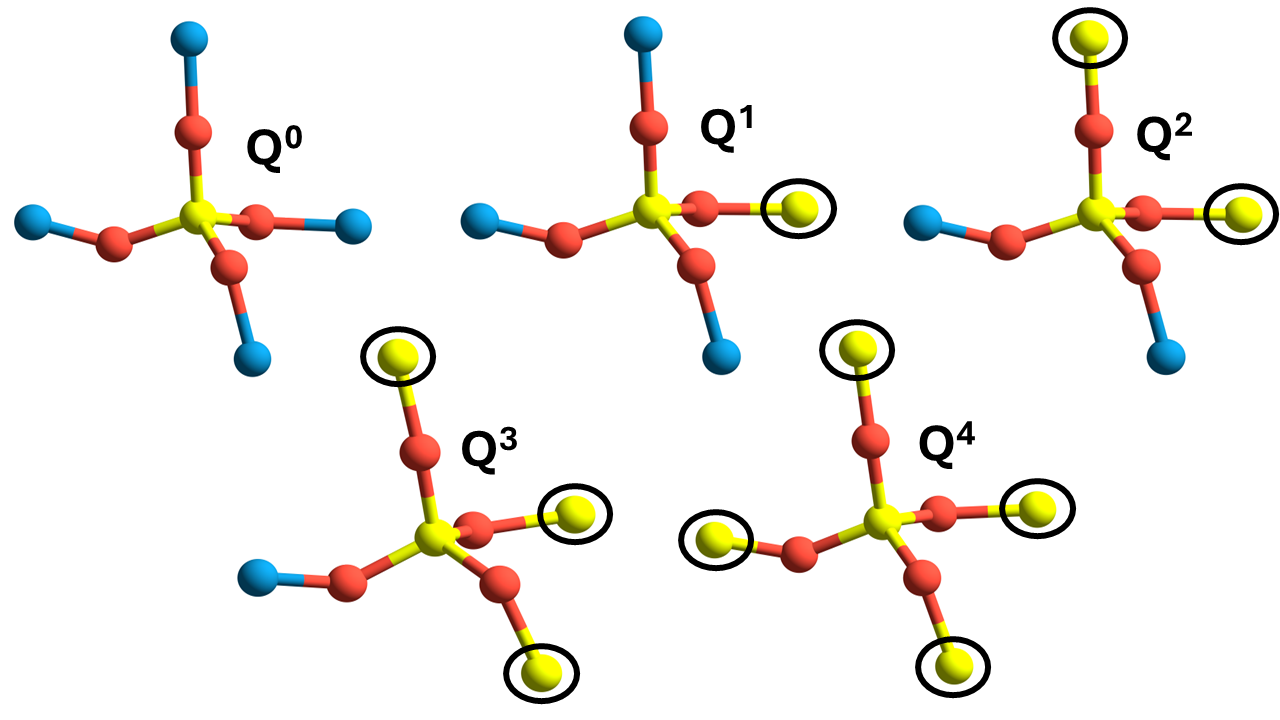}
  \caption{Examples of Si atoms belonging to different Q\textsuperscript{n} groups. The reference Si atom in each case is the Si atom in the centre of each SiO$_4$ tetrahedron. Atom colour key: Mg - blue,  Si - yellow, O - red.}
     \label{neighbours}
\end{figure}

To analyse the degree of silica-segregation in our grains, we employed the Q$^n$ notation \citep{Qn_notation_Stebbins_1987} to describe the local chemical neighbourhood of each Si atom. Here, $n$ denotes the number of -O-Si units bonded to an individual silicon atom. This classification leads to an assignment of each silicon atom in a grain to one of five groups with $n$ ranging from 0 to 4 (see Figure \ref{neighbours}). 

According to this scheme, a dominant proportion of Q\textsuperscript{0} and Q\textsuperscript{1} centres corresponds to isolated SiO$_4$ and $\equiv$Si-O-Si$\equiv$ groups, respectively, and a low degree of silica segregation. A high proportion of Q\textsuperscript{2} Si atoms indicates the presence of one dimensional silica chains or rings. Finally, Q\textsuperscript{3} and Q\textsuperscript{4} correspond to the greatest degree of silica segregation in regions where the majority of the oxygen atoms linked to a Si atom also link to another Si atom.

\begin{figure}
\centering
\includegraphics[width=\hsize]{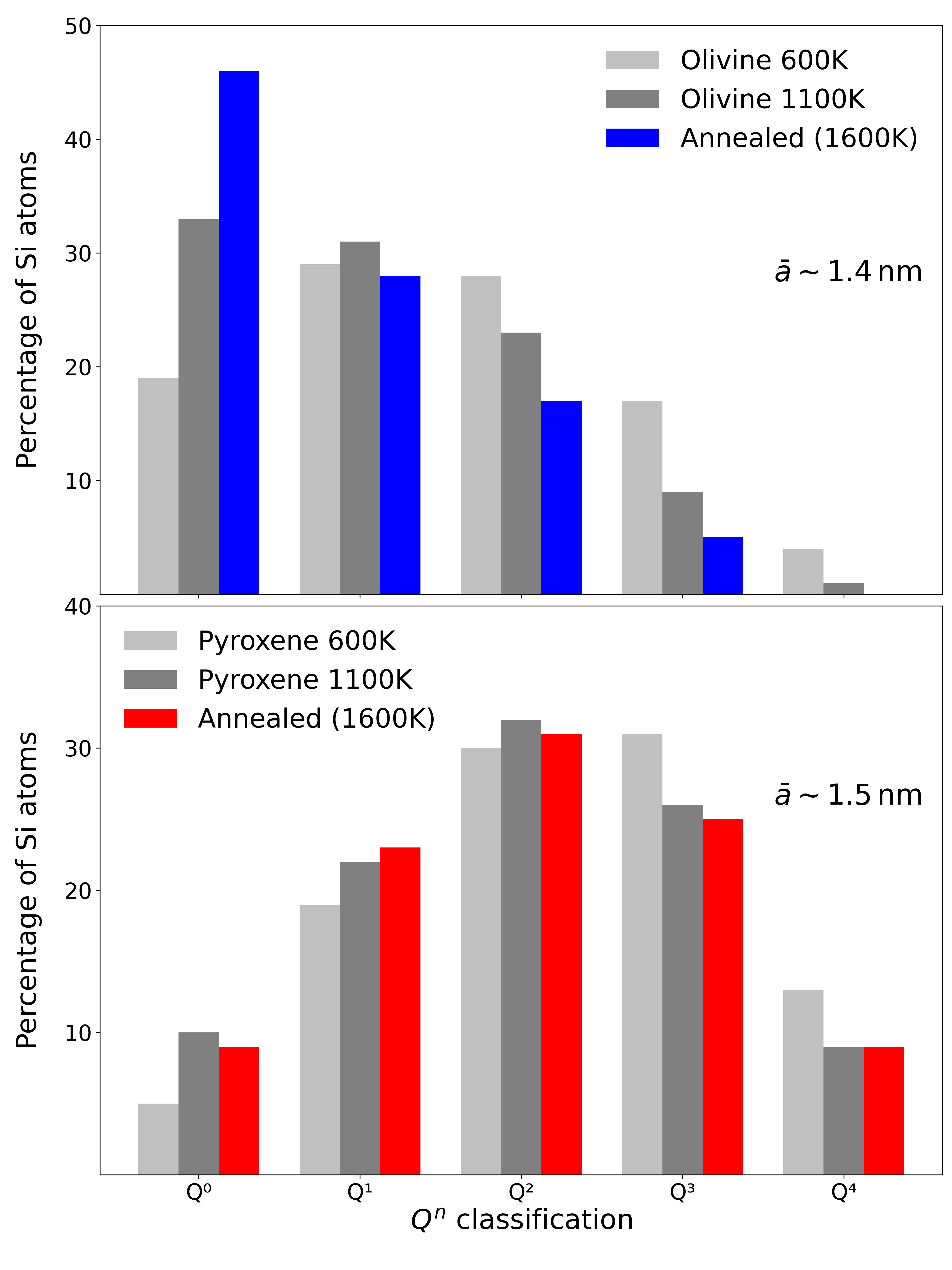}
  \caption{Top: Comparison of Q\textsuperscript{n} populations in olivinic grains formed at 600K (light grey bars) and 1000K (dark grey bars) and after annealing at 1600K (coloured bars). Bottom: Equivalent comparison for pyroxenic grains. The results correspond to averages over ten NAGGS simulations.}
     \label{segregation_annealing}
\end{figure}

The distribution of Q$^n$ centres for grains grown under the four considered scenarios are shown in Figure \ref{segregation_annealing} (see shaded light and dark gray bars). Taking the 600K scenarios first, we see a significant difference in the distribution of Q$^n$ centres with respect to the chemical composition of the corresponding nanograin. Pyroxenic grains exhibit a significant degree of silica segregation as evidenced by the high proportion of Q\textsuperscript{2} and Q\textsuperscript{3} Si centres. The presence of the former centres is consistent with the linear 1D chains of O-linked Si centres found in the bulk enstatite crystal structure. At 600K olivinic grains show less silica segregation, with most Si atoms falling into the Q\textsuperscript{1} and Q\textsuperscript{2} categories. In both cases, the distributions are broad with all types of Q\textsuperscript{n} centres having some fractional weight.

When moving to the 1100K scenarios we see that the degree of silica segregation generally reduces, i.e. most Si atoms are found in groups with smaller $n$ compared to the same type of grain grown at lower-temperature. We attribute this effect to the increased atomic mobility at higher temperatures, which allows atoms to move and occupy lower energy positions. Specifically, in olivinic grains, higher temperatures allows for a significant increase in the number of Q\textsuperscript{0} centres, while reducing the number of Q$^n$ centres with $n > 1$. This is in-line with the dominance of SiO$_4$ centres found in crystalline forsterite. For pyroxenic grains, the higher temperature growth reduced the number of Q$_n$ centres with $n > 2$, while further increasing the number of dominant Q\textsuperscript{2} centres. 

To further analyse the relationship between temperature and silica segregation, we investigated the effect of high temperature thermally annealing of various grains. This procedure involved heating the grains to 1600K over 500 ps, starting from the temperature at which the respective grain was grown. Subsequently, the temperature was held constant for 1000 ps before being decreased in a stepwise fashion by 100 K every 100 ps, with each decrease followed by a 200 ps equilibration period. This latter procedure was repeated until the original grain temperature was reached.

The resulting Q\textsuperscript{n} distributions (see blue and red bars in Figure \ref{segregation_annealing}) clearly demonstrate that high temperature thermal annealing has a more pronounced effect on olivine grains than on pyroxene grains. Compared with the 1100K growth scenario, the distribution of Q\textsuperscript{n} centres for the annealed pyroxene grains shows only minor changes. However, for the olivine grains, the annealing results in a further dramatic increase in the proportion of Q\textsuperscript{0} centres, with a decrease in the fractional weight of all other Q\textsuperscript{n} centres. Amorphous silicate grains with a forsterite composition tend to structurally anneal more easily than those with an enstatite-like composition because the structural transformation required for crystallisation is less complex. Bulk forsterite crystallises into a structure with isolated SiO$_4$ tetrahedra, which can form through relatively short-range atomic rearrangements. In contrast, bulk enstatite crystallises into a structure where SiO$_4$ tetrahedra are linked into chains, requiring more extensive polymerization and long-range ordering. This difference increases the activation energy for enstatite crystallisation compared to forsterite. We reiterate that although annealing of these nanograins tends to reduce segregation and favour local structural order, which is similar to that in the respective bulk crystalline phase, the nanoscale dimensions of these grains inhibit full crystallisation (\cite{Zamirri-NCs_2019}). 

\begin{figure}[h]
\centering
\includegraphics[width=\hsize]{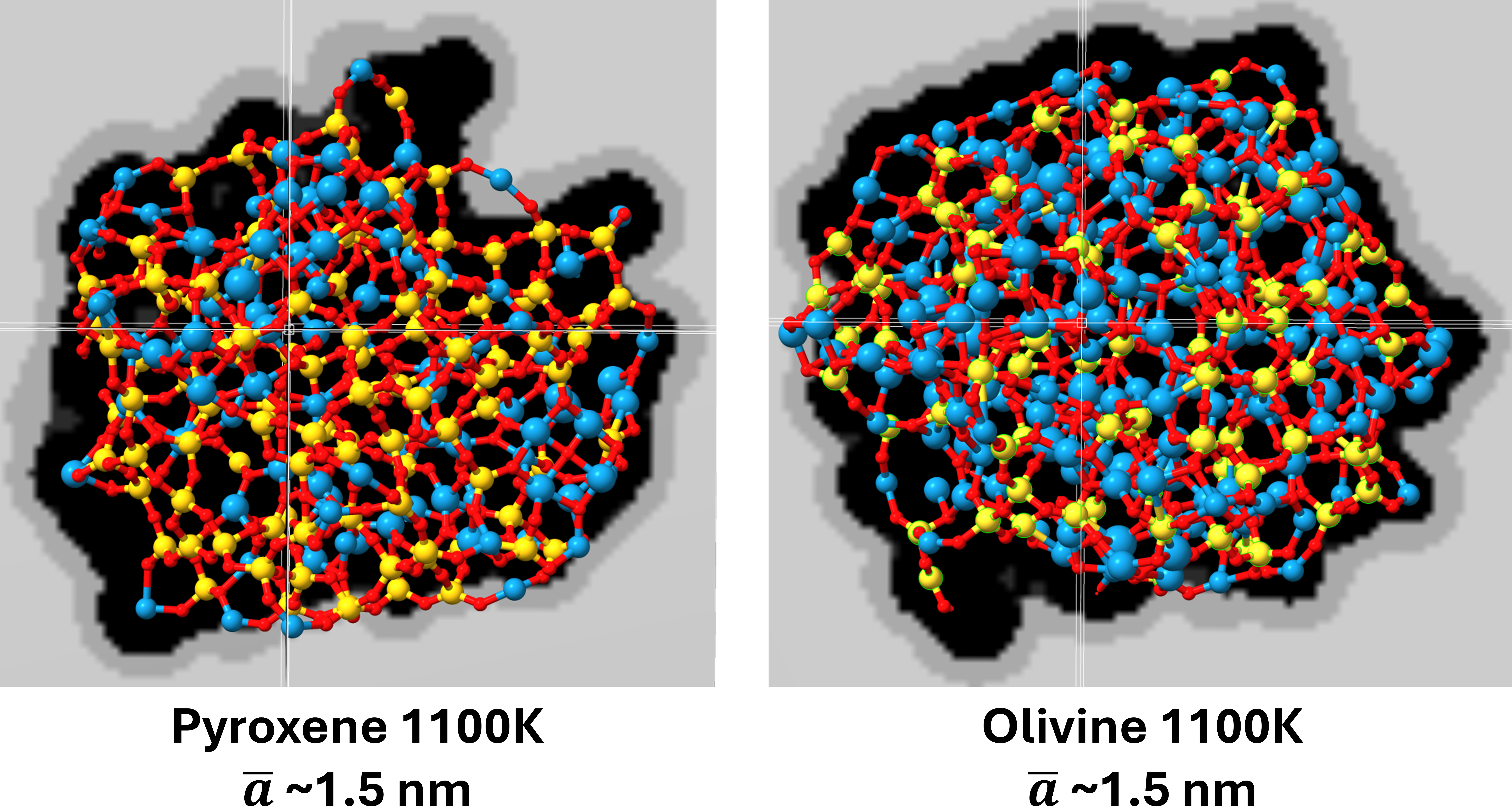}
  \caption{Visual representation of cross sections of the atomistic structure of two grains with pyroxenic composition (left) and olivinic composition (right). Both grains were obtained by NAGGS simulations at 1100K and have a $\sim$1.5 nm radii. Atom colour key: Mg - blue,  Si - yellow, O - red.}
     \label{silica_segregation}
\end{figure}

To more directly illustrate the higher degree of silica segregation in pyroxenic grains, Figure \ref{silica_segregation} shows a cross-sectional view of the atomistic structure of typical pyroxenic and olivinic grains. In the pyroxenic grain more extended (Si-O-Si)-linked regions (i.e. regions dominated by red and yellow colours) are clearly observed. Conversely, in olivinic grains, the distribution is more uniform with Mg atoms occupying more space between distributed small localised regions of Si-O bonding.

\subsection{Grain density}

\begin{figure}[h]
\centering
    \includegraphics[width=\hsize]{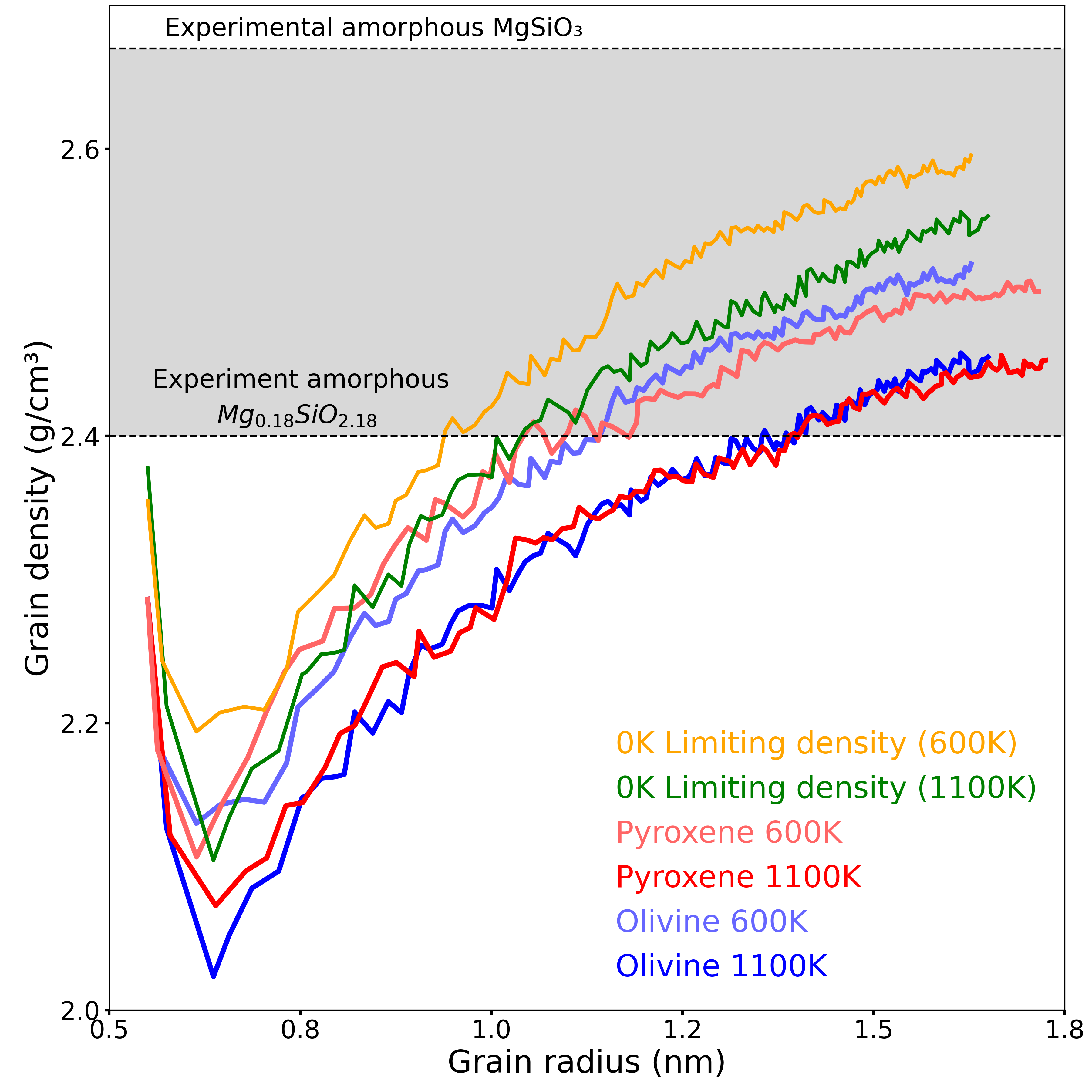}
  \caption{Evolution of grain density for pyroxenic (red) and olivinic (blue) grains at 600K and 1100K. The shaded grey  area corresponds to a range of experimentally measured densities of bulk Mg-silicate glasses reported in \cite{Jiang_2023}. Orange and green lines indicate the limiting density when optimizing the grain's structure at 0K for the grains obtained at 600K and 1100K, respectively.}
     \label{density}
\end{figure}

The density of a grain can be derived from the volume of the grain, obtained following the approach used in calculating the grain sphericity (see above), and the sum of the atomic masses. In Figure \ref{density} the evolution of grain density with respect to grain size is shown for the four considered growth scenarios.

Overall, the grain density tends to decrease with decreasing grain size. This tendency is expected due to the increasing proportion of surface area with progressively smaller grain sizes. Generally, surfaces of nanoparticles have under-coordinated atoms which bond less effectively with their local environment leading to a lower average atomic packing density.

One can also observe that the variation in density as the grains grow is relatively independent of the chemical composition for a fixed temperature. However, a higher growth temperature tends to lead to slightly lower density grains. Specifically, we see that growing grains at 1100K are about 1-1.5\% less dense than those growing at 600K. This is partly due to the higher thermal expansion of the grains at 1100K. In figure \ref{density} we also show the effect of quenching the grains to 0K while preserving their atomistic structure (i.e. to their upper limiting density relative to their specific atomistic structure). Here we see that grains quenched from 1100K still have a lower density than grains quenched from 600K, albeit with a slightly reduced density difference relative to the pre-quenched grains. This implies that the thermal expansion is not the only reason for the temperature-dependent density differences.   

Another characteristic feature in all the density plots in figure \ref{density} is an initial drop in density in the first few growth steps. This decrease is due to the higher susceptibility of smaller grains to the impacts of the incoming monomers. These impacts cause structural distortions in the grain when it is very small, which lead to a small expansion of its volume. This effect is more pronounced for higher-temperature simulations due to the higher velocity monomer impacts, leading to a greater drop in density. It appears that this effect produces an initial offset in the density of the growing grains that then persists throughout the growth.    

For a more quantitative measure of our calculated densities, in figure \ref{density} we also compare our values with experimentally measured densities for magnesium silicate glasses (\cite{Jiang_2023}). For smaller grains ($< 1$ nm radius) the densities of our growing amorphous grains are found to be lower than that of bulk amorphous Mg-silicates. However, for densities of the largest grains tend to converge to values within the measured range typical for bulk glassy Mg-silicates. We note that the experimental values were likely obtained at room temperature and will thus exhibit very little thermal expansion.

We also note that the density of porous aggregates of (nano)silicate grains would have much lower densities than those reported here. In future versions of the NAGGS code we plan to incorporate grain-grain aggregation to explore the formation and properties of nanoporous grains \citep{Porous_grain_review}. 

\subsection{Electric dipole moments}

Nanosized grains strongly contribute to photoelectric heating of the ISM (\cite{Draine_photoelectric_2001}) and the intergalactic medium (\cite{Nath_photoelectric_1999}) and are also involved in regulating grain-plasma coupling through dust charging (\cite{Dust_plasma_interactions_review_2009}). These processes are intimately linked with the electric dipole of moments of the grains. The tendency for nanograins to aggregate can also be influenced by dipole-dipole interactions (\cite{Dipole-dipole_interactions_2009}). The dipole moment of a nanograin is also a crucial parameter for assessing its potential to emit in the microwave region of the electromagnetic spectrum. In this context, nanosilicate grains have been proposed as potential carriers of the anomalous microwave emission (AME) ((\cite{AME_no_PAHS_Hensley_2017}, \cite{Hoang_2016}, \cite{ame_toni}). 

The AME is a foreground feature, with a broad peak around 20-30 GHz, and is observed in various astrophysical environments \citep{ame}. For nanosilicates to contribute to the AME they must be small enough to spin fast enough at typical temperatures encountered in the ISM (<300K). Ultrasmall nanosilicates with <100 atoms are readily excited to sufficient rotational velocities. DFT calculations have also directly shown that they also have high enough dipoles to explain the AME with only 1\% of the available Si mass in such grains \citep{ame_toni}. The larger nanosilicate grains considered herein, are expected to have higher dipole moments, but will spin at a slower rate. However, when taking into account the complex rotational dynamics of non-spherical grains (e.g. wobbling effects \cite{Wobbling_Hoang_2010}) and the alignment of the dipole moment with the inertial axes (\cite{Dipole_alignment_Hoang_2011}) the microwave emission peak of such grains can be shifted to the regime of the AME peak.        

One of the most important parameters that determines the potential for a nanosized grain to be a potential carrier of the AME is the size-normalised dipole moment ($\beta$),

\begin{equation}\label{beta_equation}
    \beta = \frac{\mu}{\sqrt{N}},
\end{equation}

where $\mu$ is the total dipole moment of the grain in units of Debye and $N$ is the number of atoms in the grain. In most numerical modelling of spinning nanograins, $\beta$ is taken to be a constant that characterises the dipolar character of the grain material. The electric dipole magnitude then scales as $\beta\sqrt{N}$. In our atomistic FF-based treatment, $\mu$ can be directly calculated from 

\begin{equation}\label{dipole_equation}
    \mu = \sum_i \bm{r_i}\bm{q_i} + Q\bm{R_{COM}},
\end{equation}

where the $\bm{r_i}$ and $\bm{q_i}$ are the positions and the ionic and shell charges (see FF parameters in Appendix A), $Q$ is the net grain charge, and $\bm{R_{COM}}$ is the position of the COM. To confirm the reliability of the FF for calculating accurate dipole moments we compare the predictions of the FF with those from DFT calculations for a set of nanosilicates (see Appendix C). The excellent correlation between the two sets of evaluated data show that our employed FF tends to slightly underestimate nanosilicate dipole moments by $\sim$5\%. The dipole moments used in the following are thus adjusted by 5\% to correct for this small systematic shift.

\begin{figure}[h]
\centering
\includegraphics[width=\hsize]{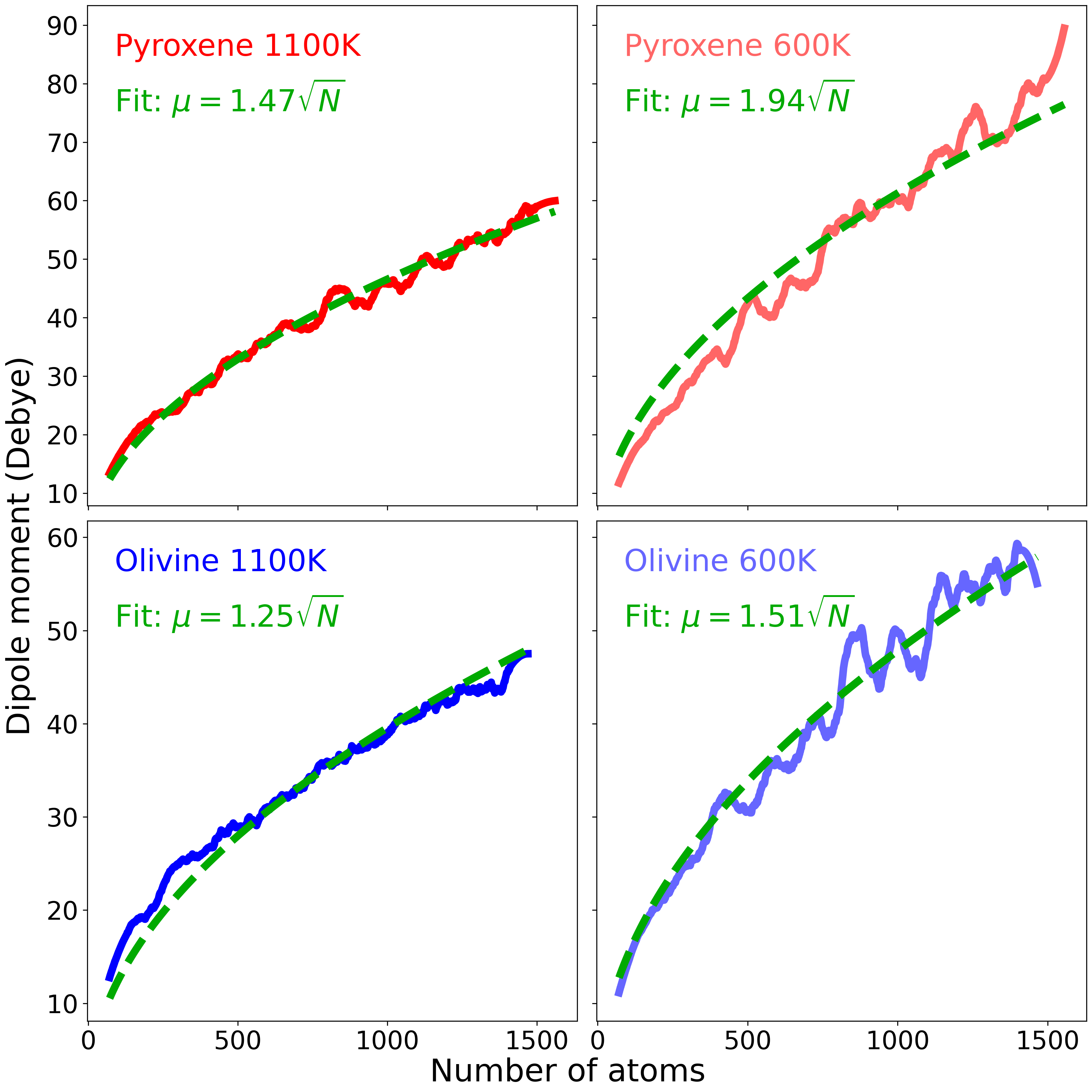}
  \caption{Evolution of the grains dipole moment for the simulation of pyroxenic composition at 1100K (top left), and 600K (top right), olivinic composition at 1100K (bottom left) and 600K (bottom right). The dipole moments have been corrected to have DFT-like accuracy according to \ref{dipoles_Comparison}. The results in each case were obtained as the average over ten NAGGS runs.}
     \label{mu_dipoles}
\end{figure}

In figure \ref{mu_dipoles} we show the evolution of $\mu$ with respect to grain size for each of the four growth scenarios considered. For each scenario we show a plot of $\beta\sqrt N$ with a fitted constant value of $\beta$ to best match the observed dipole growth rate. All fitted $\beta$ curves provide a reasonable description of the calculated size-dependent dipole evolution. However, we find that pyroxene grains tend to have higher $\beta$ values than olivine grains (by 18-28 \%) for the same growth temperature. This shows that $\beta$ is not constant for a particular silicate material. We also see that higher temperature growth leads to a significantly lower respective $\beta$ value for the same material. This behaviour could be attributed to the increased mobility of the atoms at higher temperatures, allowing them to find more stable positions, which tends to reduce the total dipole moment of the grain. 

In all grain growth scenarios we find that $\beta > 1$ . In numerical modelling studies of the potential contribution of nanosilicates to the AME, a maximum upper bound of $\beta =1.0$ is typically assumed (\cite{Hoang_2016, AME_no_PAHS_Hensley_2017}). We note that the higher $\beta$ values we find are in line with those from accurate quantum chemical DFT calculations for ultrasmall nanosilicate grains \citep{ame_toni}. Generally, amorphous silicates exhibit local structural ordering based on alternating anionic (but non-dipolar) SiO$_4$$^{4-}$ groups and Mg$^{2+}$ cations. The cumulative effect of these local charge imbalances produces substantial net dipole moments, as confirmed by all our nanosilicate models. This characteristic may may have important implications for understanding their potential role as carriers of anomalous microwave emission.

\begin{figure}[h]
\centering
\includegraphics[width=\hsize]{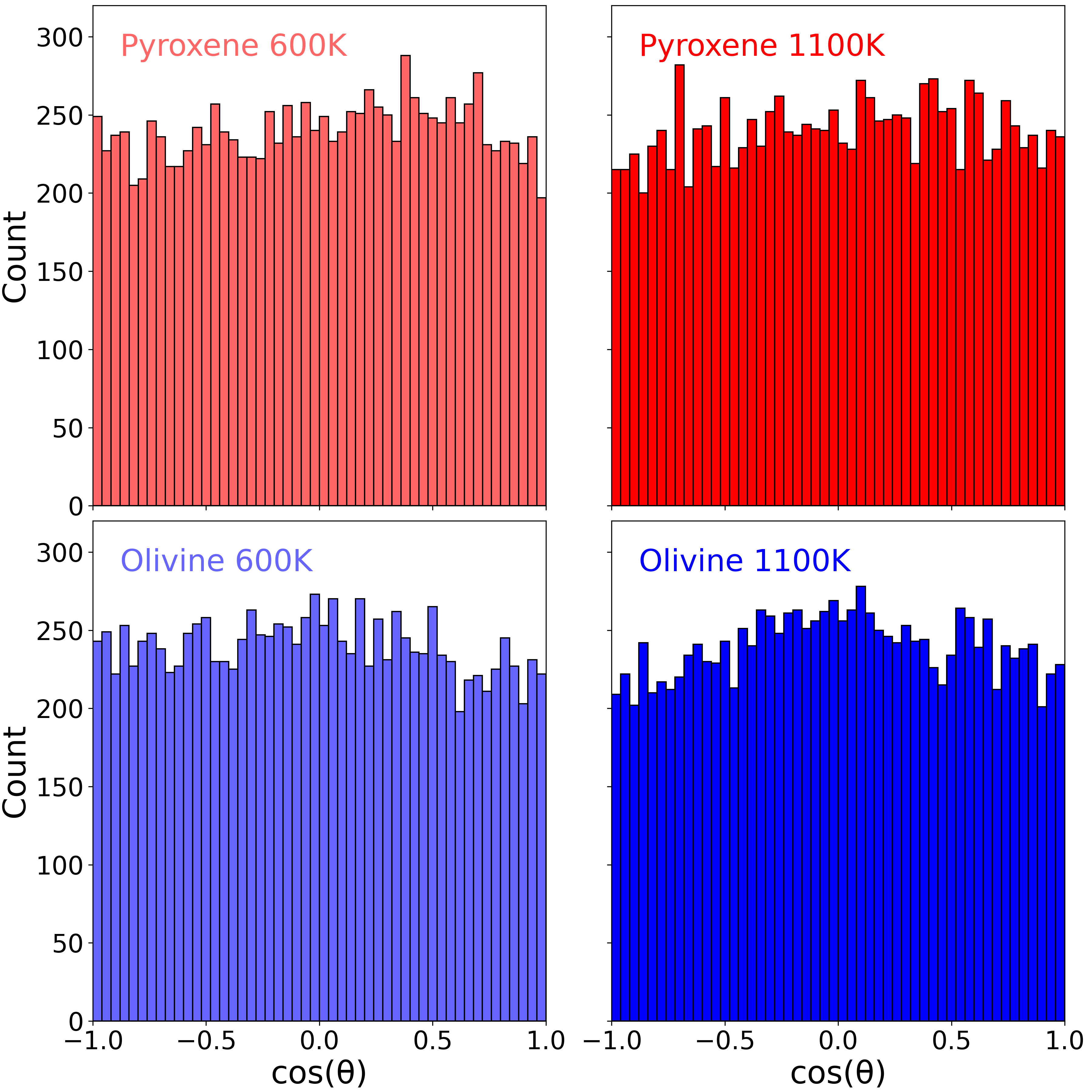}
  \caption{Histograms of the cosine of the angle between each grain’s dipole moment and its principal (maximum) moment of inertia for simulations of pyroxenic (top, red) and olivinic (bottom, blue) grains at 600 K (left) and 1100 K (right). Each histogram combines data from all grains obtained across ten  NAGGS runs.}
     \label{alignment_dipoles}
\end{figure}

As the nanosilicate grains in our simulations do not have perfectly spherical shapes (see above) we have also calculated the alignment of the dipole moment with the largest axis of inertia (i.e. around which grain rotation tends to be favoured with energy dissipation). In figure \ref{alignment_dipoles} we show histograms of the cosine of this alignment angle ($\theta$) considering all cycles of each grain growth scenario. Here, for all cases, we see that the distributions are very flat, which indicates that there is no favoured alignment between the dipole moment and the principal inertial axes. In such situations, grains can exhibit complex rotational behaviour when interacting with electromagnetic fields that alters their alignment, emission, and charging characteristics. In turn, this can lead to significant consequences for interpreting polarization and dust-plasma interactions in astrophysical environments. 

\section{Conclusions}
We have demonstrated how NAGGS can provide realistic atomistically detailed nanograin models by simulating the dynamic process of growth by monomeric accretion at finite temperatures. Using the circumstellar growth of Mg-rich nanosilicates as an example case study, we focused on two different chemical compositions and two limiting growth temperatures. Our conclusions are based on an analysis of a set of astrophysically relevant properties of the produced grain models with respect to size, chemical composition, and growth temperature.          
  \begin{enumerate}
      \item All growing nanosilicate grains, regardless of composition or growth temperature, are amorphous and have a slight tendency to be prolate. All grains become more spherical with increasing size.
      \item Olivine and pyroxene nanograins grown at the same temperature have similar densities. Grains grown at higher temperatures tend to have lower densities. We mainly attribute this effect to the influence of high temperature accretion on the initial seed.   
      \item For the same grain size, pyroxene grains have larger specific surface areas than olivine grains. This difference is due to the higher roughness of the pyroxene grain surfaces. In turn, this is likely linked to the high degree of silica segregation in the pyroxene grains. 
      \item Increasing the growth temperature from 600K to 1100K tends to lead to less locally segregated nanosilicate structures. This effect is more pronounced for olivinic nanograins. High temperature annealing at 1600K has little further effect on the segregation of pyroxene nanograins but leads to a further significant lowering of segregation in the olivine nanograins.    
      \item The size-normalized electric dipole moments of the nanosilicate grains are consistently high ($\beta > 1$ in all cases), indicating that these grains can efficiently produce rotational emission and may therefore play a significant role as carriers of AME.
      \item Pyroxene nanograins exhibit systematically higher dipole moments than olivine nanograins of comparable size. For both compositions, increasing the growth temperature results in a modest decrease in their corresponding $\beta$ values.
      \item For all considered grain growth scenarios, there is no statistically significant alignment of the dipole moment of the grain with any of its principal inertial axes. As the dipole orientation is effectively random relative to the rotation axis, it could potentially further enhance microwave emission.  
   \end{enumerate}
We note that these insights are only possible due to: i) the simulation of grain grown by monomer accretion to produce realistic atomistically detailed grain models, and ii) the ability to directly interrogate the structures and properties of the resulting grain models. We anticipate that such an approach will help improve our understanding of how the properties of nanograins lead to observational consequences. We thus hope that NAGGS will provide a tool for placing new constraints on the abundance and character of nanosilicates and other nanosized grains in a range of astrophysical environments. 

\begin{acknowledgements}
      S.T.B. and J.M.G. acknowledge financial support from the Spanish Ministerio de Ciencia, Innnovación y Universidades (PID2024-157971NB-C22,PID2021-127957NB-I00, TED2021-132550B-C21 and CEX2021-001202-M via the Spanish Structures of Excellence María de Maeztu program) and the Generalitat de Catalunya (2021SGR00354). J.M.G. also acknowledges the Generalitat de Catalunya for the pre-doctoral grant 2020 FI-B-00617.
\end{acknowledgements}

\bibliographystyle{aa}
\bibliography{Biblio}

\begin{appendix} 

\section{FF parameters used in the MD simulations} \label{IP_Parameters}

The interactions between ions are described by an empirical FF based on pairwise interactions described by a Buckingham potential and Coulombic interactions:
\[
U_{ij}(r)=\frac{q_i q_j}{4\pi\varepsilon_0\,r}
+ A_{ij}\,\mathrm{e}^{-r/\rho_{ij}} - \frac{C_{ij}}{r^{6}} \,,
\]
where $U_{ij}(r)$ is the potential energy of two interacting ions $i$ and $j$ separated by a distance $r$. The values of all constants are provided below.
\begin{table}[H]
\caption{Ionic charges}            
\label{IP_Charges}      
\centering                          
\begin{tabular}{c c }        
\hline\hline                 
Element & Charge (q) \\    
\hline                        
   Si & 2.7226 \\      
   Mg & 1.3613 \\
   O\textsubscript{Core} & 1.91981 \\
   O\textsubscript{Shell} & -3.28111
\end{tabular}
\end{table}

\begin{table}[h]
\caption{Buckingham parameters}             
\label{IP_Buckingham}      
\centering                         
\begin{tabular}{c c c c}        
\hline\hline                 
Interacting species & A(eV) & $\rho$ (\AA) & C (eV $\AA^6$)  \\    
\hline                        
   Si-O\textsubscript{Shell} & 8166.2632 & 0.193884 & 0.0 \\     
   O\textsubscript{Shell}-O\textsubscript{Shell} & 15039.909 & 0.227708 & 0.0 \\
   Mg-O\textsubscript{Shell} & 2014.61 & 0.247732 & 10.1938
\end{tabular}
\end{table}
In addition to the interionic interactions, polarisation of the oxygen anions is described by the shell model with a harmonic spring between a positive core and a massless negative shell (\cite{FF_shell_model}). 
\begin{table}[h]
\caption{Core-Shell spring constant}             
\label{IP_Core-Shell}     
\centering                         
\begin{tabular}{c c}        
\hline\hline                 
Interacting species & Spring constant ($\AA^{-2}$)  \\
\hline                        
   O\textsubscript{Core}-O\textsubscript{Shell} & 256.71027    
\end{tabular}
\end{table}

We note that this the parameterisation of this FF is very similar to that used in \cite{Escatllar_2019}.
\newpage
\section{Comparison of energetic stabilities of a set of 90 ultrasmall nanosilicate structures as calculated using the employed FF and DFT calculations}

\begin{figure}[h]
\centering
\includegraphics[width=\hsize]{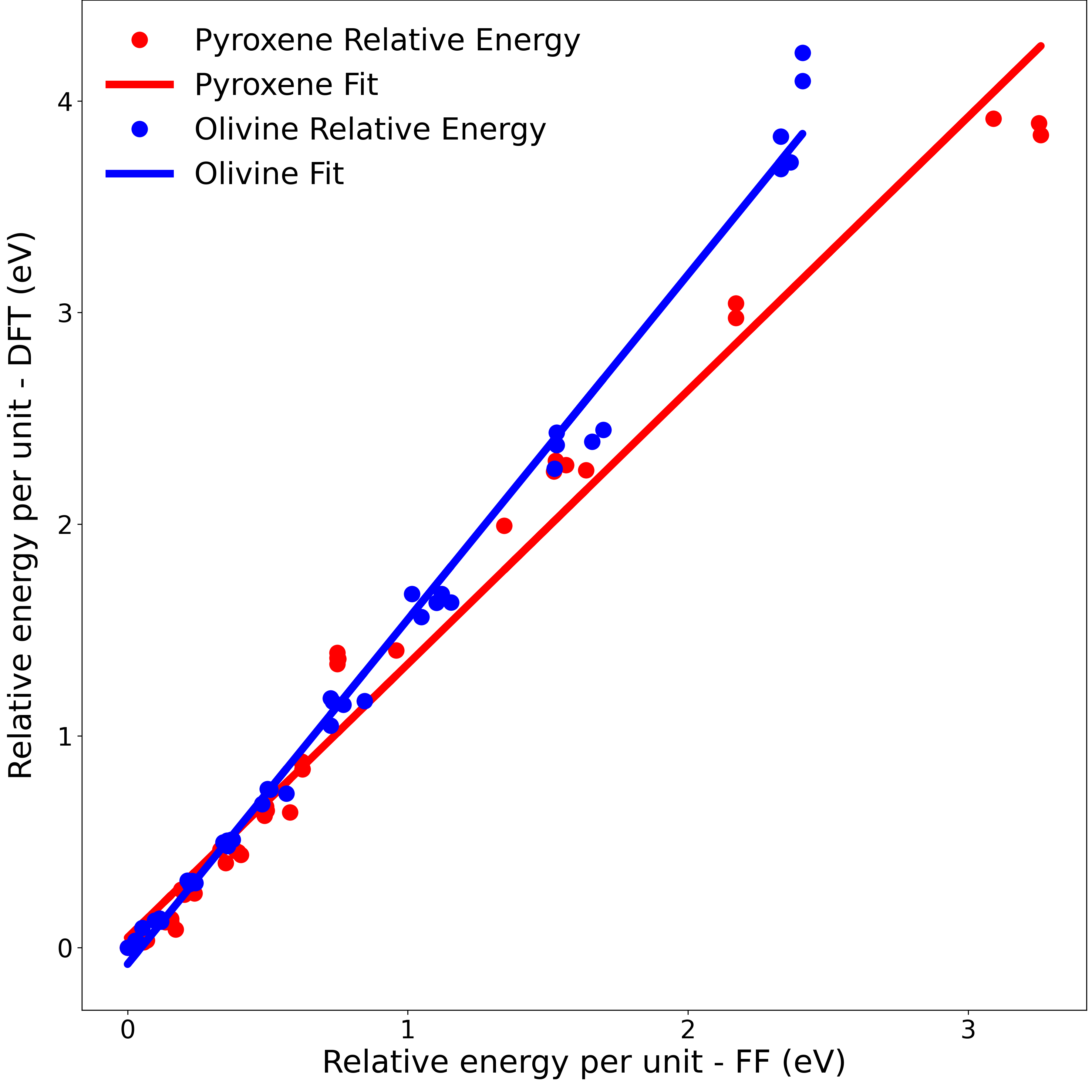}
  \caption{Comparison of the relative energies of the optimized structures of all nanosilicates in the  set used in \cite{FaradayDiscuss_paper}, as calculated using DFT and the FF used in the NAGGS simulations. The DFT calculations employed the PBE0 functional (\cite{pbe0}) and a Tier1/tight basis set based on all-electron numeric atom-centered orbitals (\cite{NAO_basis_sets_elephant}), as implemented in the FHI-AIMS code (\cite{FHI-AIMS}).}
    \label{dft_ff_relative_energies}
\end{figure}
\newpage
\section{Comparison of dipole moments of a set of 90 ultrasmall nanosilicate structures as calculated using the employed FF and DFT calculations}

\begin{figure}[H]
\centering
\includegraphics[width=\hsize]{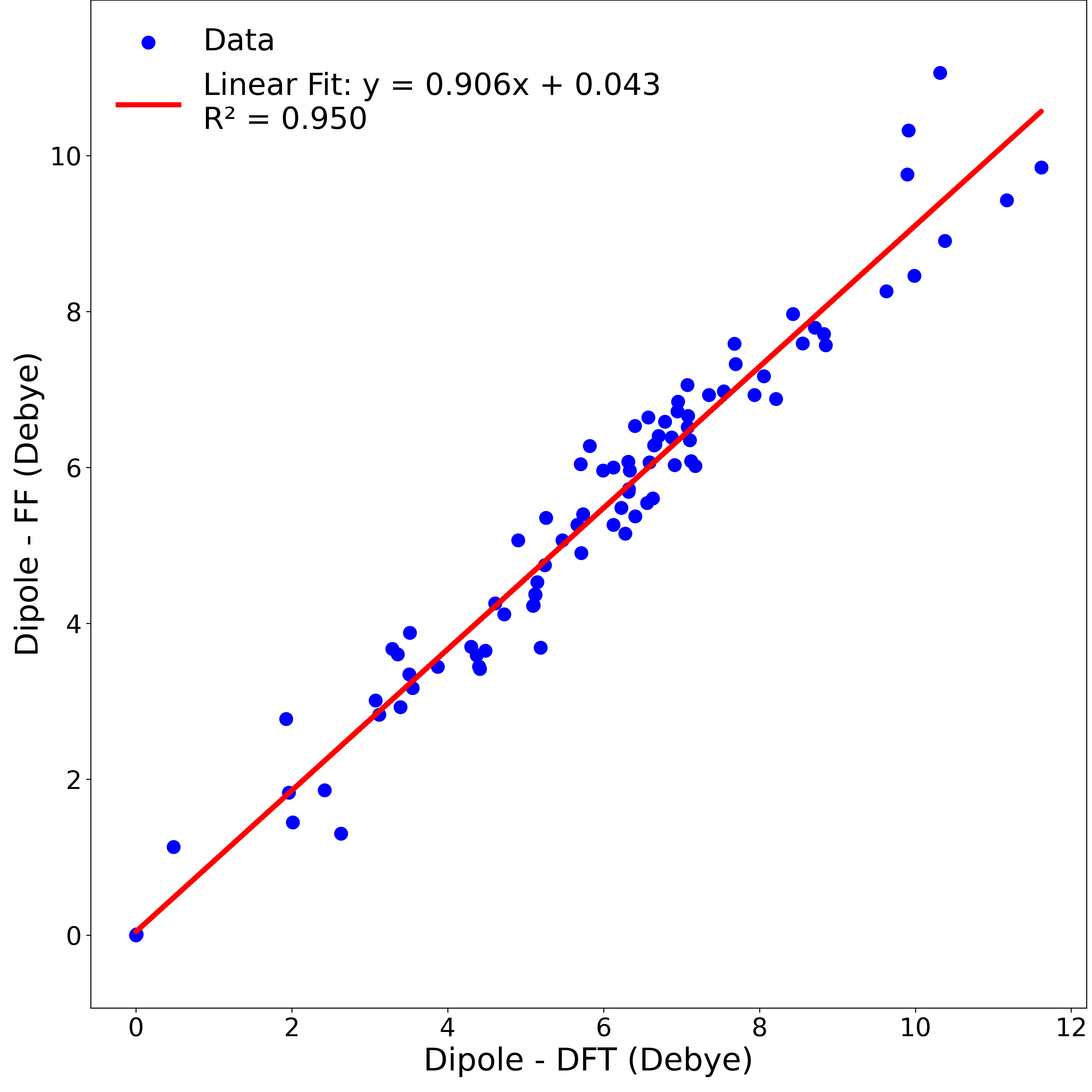}
  \caption{Magnitudes of dipole moments of the optimized structures of all nanosilicate used in \cite{FaradayDiscuss_paper}, using DFT and the FF used in the NAGGS simulations. DFT calculations used the same set-up as in used in Appendix B. We note that DFT calculations employing the PBE0 functional are known to provide highly accurate dipole moments \cite{PBE0_dipoles}.}
     \label{dipoles_Comparison}
\end{figure}
\newpage

\section{Evolution of the O:Si ratio} \label{O_Si_ratio}

\begin{figure}[H]
\centering
\includegraphics[width=\hsize]{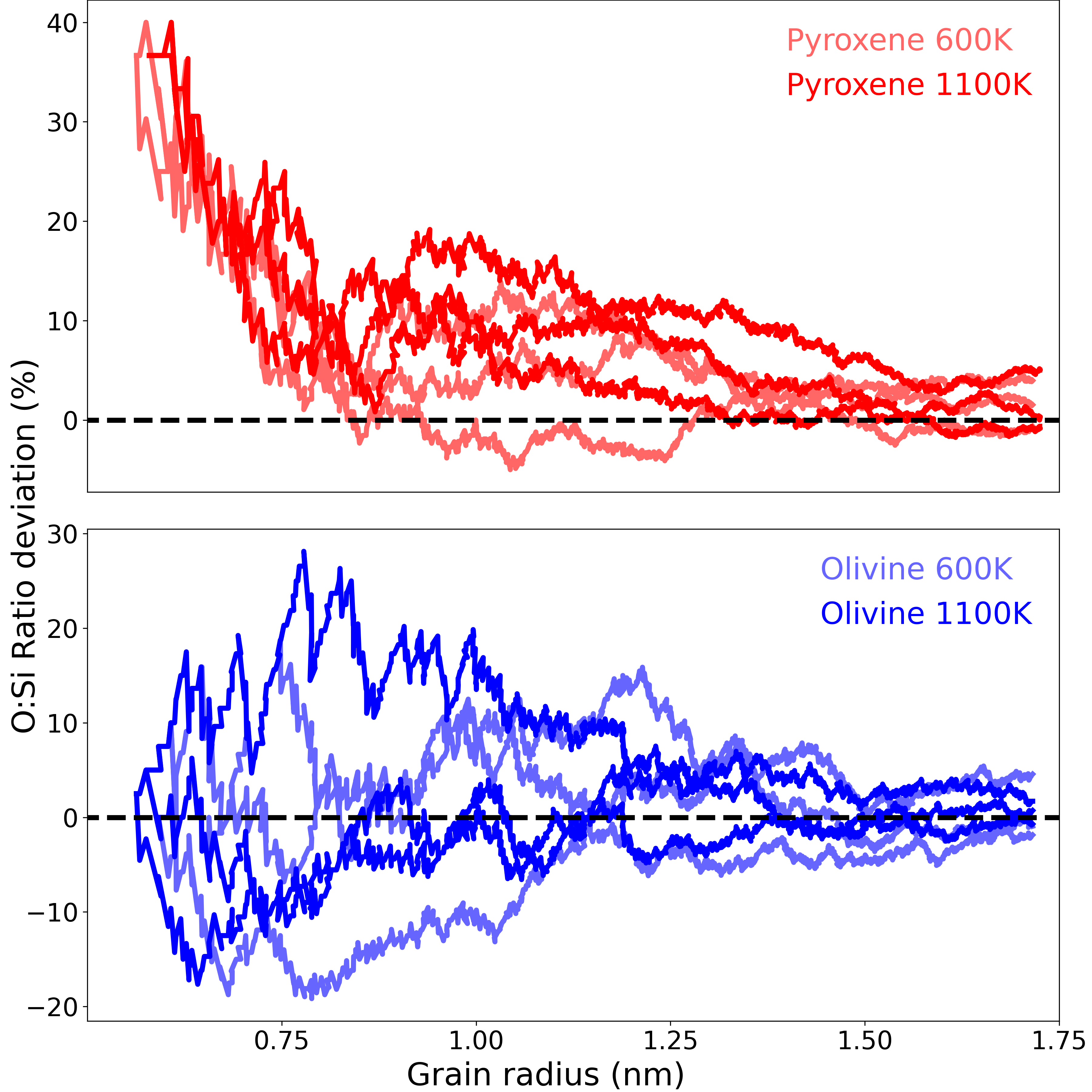}
  \caption{Size-dependent convergence of the O:Si ratio for pyroxenic (top) and olivinic (bottom) grain compositions, grown at 600 K and 1100 K. For each chemical composition and temperature, three different simulations are shown for each composition-temperature combination.}
     \label{o_si_ratio_plot}
\end{figure}

\section{Radius evolution for simulations carried out at 600K} \label{SI_radius_600K}

\begin{figure}[H]
\centering
\includegraphics[width=\hsize]{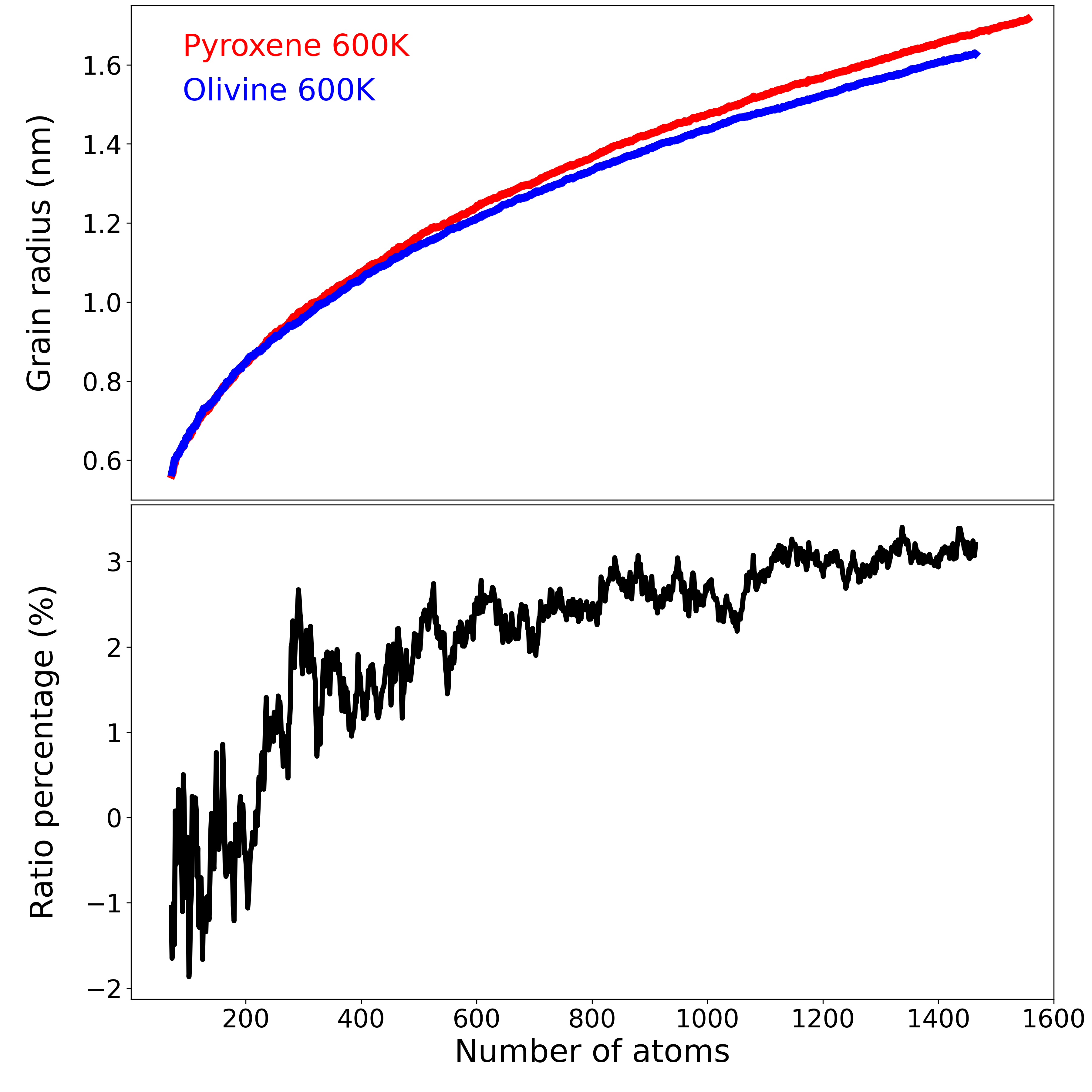}
  \caption{Top: Evolution of $\bar{a}$ of pyroxene (red) and olivine (blue) grains grown at 600K with respect to number of atoms. Bottom: the ratio of the radii of pyroxene and olivine grains. Results are obtained from averages over ten NAGGS simulations.}
     \label{radius_600K}
\end{figure}

\section{Sphericity evolution for simulations carried out at 600K} \label{SI_sphericity_600K}

\begin{figure}[H]
\centering
\includegraphics[width=\hsize]{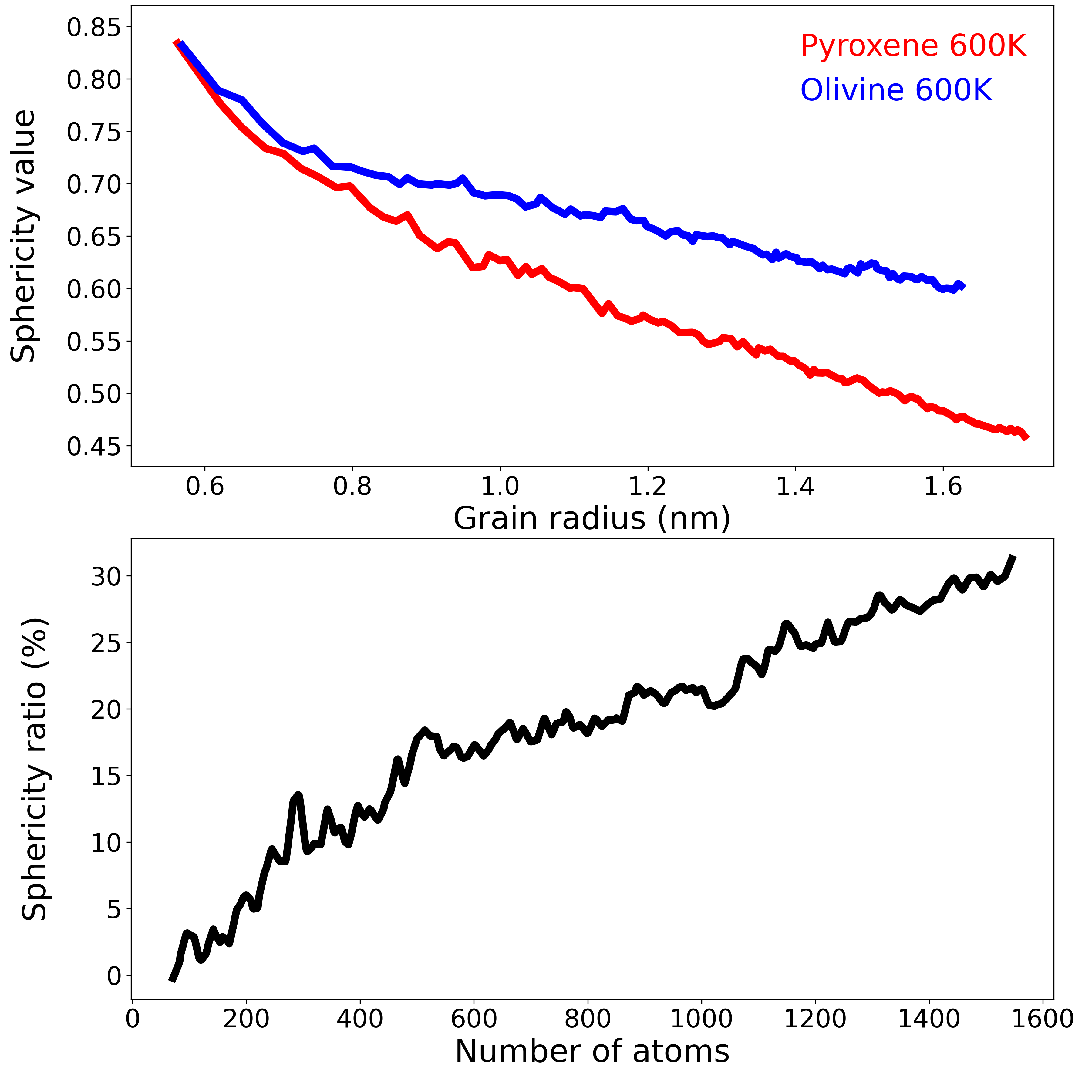}
  \caption{Top: Sphericity with respect to the number of atoms in the grain for  pyroxene (red) and olivine (blue) grains obtained at 600K. Bottom:  ratio of pyroxene to olivine grain sphericity. The results are obtained from averages over ten NAGGS simulations.}
     \label{sphericity_600K}
\end{figure}
\newpage
\section{Surface increase evolution for simulations carried out at 600K} \label{SI_surface_600K}

\begin{figure}[H]
\centering
\includegraphics[width=\hsize]{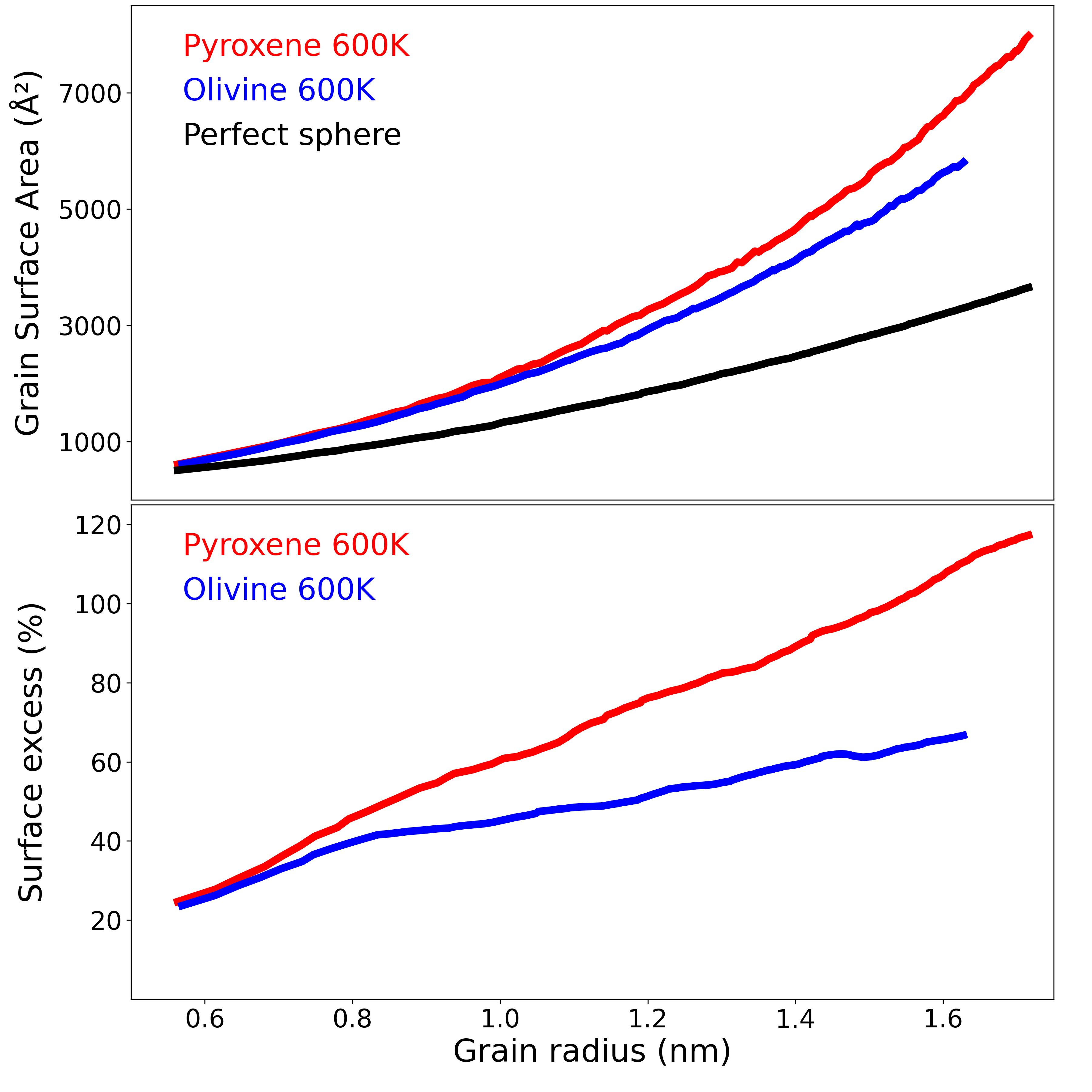}
  \caption{Top: Surface area of growing pyroxenic (red) and olivine (blue) grains grown at 600K, and of a sphere with same radius as the grain (black). Bottom: Percentage of excess surface area of the grains with respect to the corresponding sphere. The results are obtained from averages over ten NAGGS simulations.}
     \label{surface_600K}
\end{figure}

\end{appendix}

\end{document}